\documentclass[sigconf]{acmart}

\usepackage{graphicx}
\usepackage{microtype}
\usepackage{subcaption}
\usepackage{booktabs} 
\usepackage{dsfont}
\usepackage{enumitem}
\usepackage{multirow} 
\usepackage{makecell}
\usepackage{booktabs}
\usepackage{pifont}
\usepackage{tikz}
\usepackage[table]{xcolor}
\newcommand{\cmark}{\ding{51}} 
\newcommand{\xmark}{\ding{55}} 
\usepackage{hyperref}
\usepackage{algorithm}
\usepackage{algorithmic}
\usepackage{amsmath}
\usepackage{amssymb}
\usepackage{mathtools}
\usepackage{amsthm}

\AtBeginDocument{%
  }

\copyrightyear{2026}
\acmYear{2026}
\setcopyright{cc}
\setcctype{by}
\acmConference[MM '26] {Proceedings of the 34th ACM International Conference on Multimedia}{November 10--14, 2026}{Rio de Janeiro, Brazil}
\acmBooktitle{Proceedings of the 34th ACM International Conference on Multimedia (MM '26), November 10--14, 2026, Rio de Janeiro, Brazil}
\acmISBN{979-8-4007-2213-4/2026/11}
\acmDOI{10.1145/3767308.3836139}

\begin{document}

\title{Noise as a Probe: Membership Inference Attacks on Diffusion Models Leveraging Initial Noise}


\author{Puwei Lian}
\email{220255555@seu.edu.cn}
\orcid{1234-5678-9012}
\affiliation{%
  \institution{Southeast University}
  \city{Nanjing}
  \country{China}
}

\author{Yujun Cai}
\email{yujun.cai@uq.edu.au}
\affiliation{%
  \institution{The University of Queensland}
  \city{Brisbane}
  \country{Australia}
}

\author{Songze Li}
\authornote{Corresponding author.}
\email{songzeli@seu.edu.cn}
\affiliation{%
  \institution{Southeast University}
  \city{Nanjing}
  \country{China}
}

\affiliation{%
  \institution{Engineering Research Center of Blockchain Application, Supervision and Management (Southeast University), Ministry of Education}
  \city{Nanjing}
  \country{China}
}

\author{Bingkun Bao}
\email{bingkunbao@hfut.edu.cn}
\affiliation{%
 \institution{Hefei University of Technology}
 \city{Heifei}
 \country{China}
}


\renewcommand {\shortauthors}{Puwei Lian, Yujun Cai, Songze Li, and Bingkun Bao}
\begin{abstract}
Diffusion models have achieved remarkable progress in image generation, but their increasing deployment raises serious concerns about privacy and copyright. In particular, fine-tuned models are highly vulnerable, as they are often fine-tuned on small and private datasets. Membership inference attacks (MIAs) are used to assess privacy risks by determining whether a specific sample was part of a model’s training data. Existing MIAs against diffusion models either assume obtaining the intermediate results or require training the shadow model on auxiliary datasets. In this work, we utilized a critical yet overlooked vulnerability: the widely used noise schedules fail to fully eliminate semantic information in the images, resulting in residual semantic signals even at the maximum noise step. We empirically demonstrate that the fine-tuned diffusion model captures hidden correlations between the residual semantics in initial noise and the original images. Building on this insight, we propose a simple yet effective membership inference attack, which injects semantic information into the initial noise and infers membership by analyzing the model’s generation result. Extensive experiments demonstrate that the semantic initial noise can strongly reveal membership information, highlighting the vulnerability of diffusion models to MIAs. Code is available at \url{https://github.com/S3IC-Lab/NoiseMIA}.
\end{abstract}




\maketitle

\section{Introduction}
\label{introduction}
Diffusion models have shown outstanding performance in generating high-quality images. With the widespread release of large-scale pre-trained models, users can easily download and fine-tune them on downstream datasets. However, this convenience raises serious concerns about the copyright and privacy of training data \cite{wen2024detecting, ren2024unveiling}. Membership inference attacks (MIAs) are crucial methods for assessing the privacy risks. These attacks typically exploit the model's overfitting behavior, leveraging the differences in its responses to training versus non-training samples to infer membership status \cite{yeom2018privacy}. In the context of diffusion models, MIAs aim to determine whether an image was used for training \cite{MIA}. In particular, fine-tuning is widely regarded as the stage most prone to privacy leakage, since the datasets used are relatively small and often private (e.g., personal photos, proprietary artwork) \cite{li2024shake, pang2023black}. Consequently, studying MIAs on fine-tuned diffusion models is significant for understanding and mitigating potential privacy risks.

\begin{figure}[t]
    \centering
    \setlength{\tabcolsep}{1pt} 
    \renewcommand{\arraystretch}{0.5} 
     \begin{tabular}{cccc}
        \multicolumn{2}{c}{Member} & \multicolumn{2}{c}{Non-Member} \\[2pt]
        \cmidrule(lr){0-1} \cmidrule(lr){3-4}
        Original & Generated & Original & Generated \\
        \includegraphics[width=0.24\linewidth]{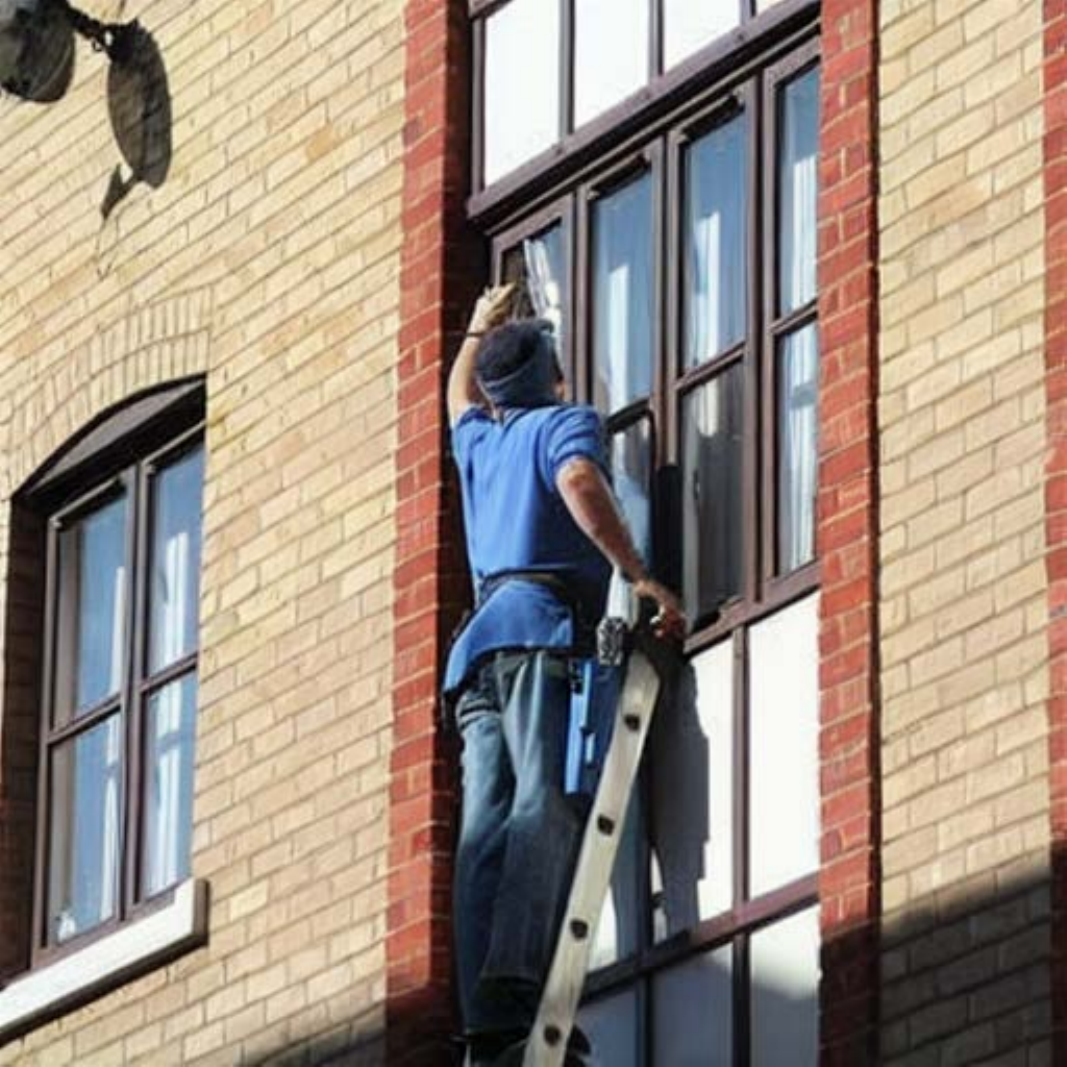} &
        \includegraphics[width=0.24\linewidth]{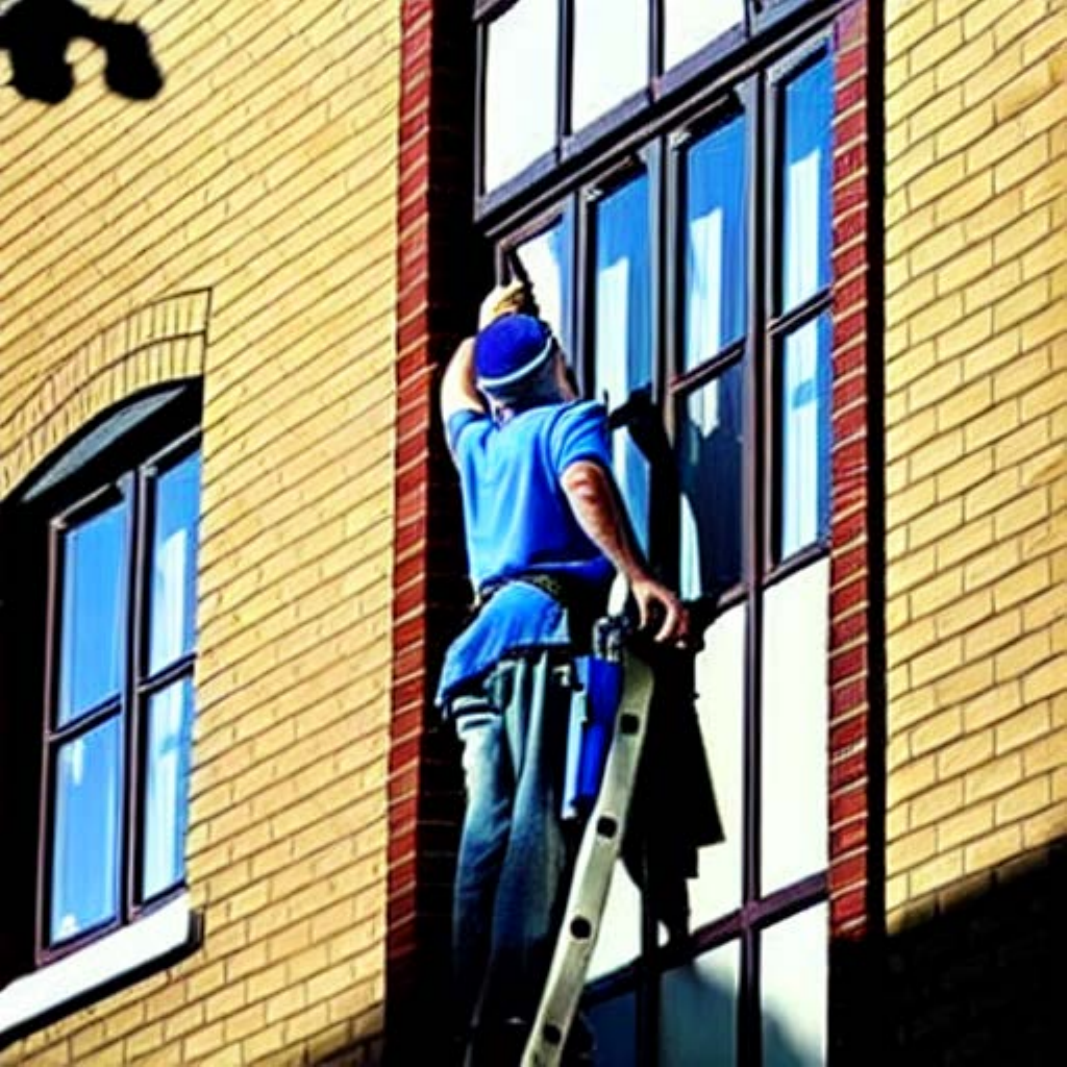} &
        \includegraphics[width=0.24\linewidth]{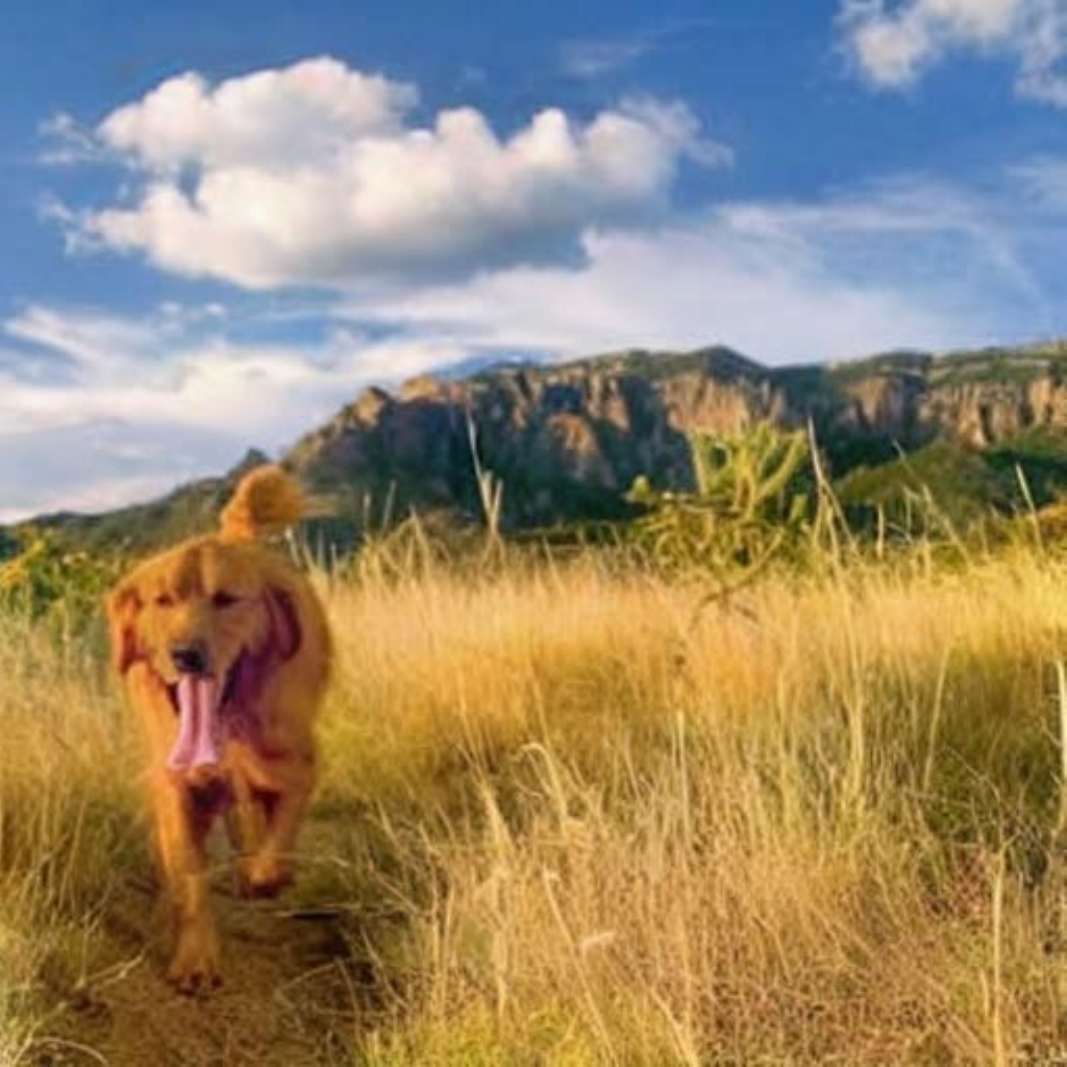} &
        \includegraphics[width=0.24\linewidth]{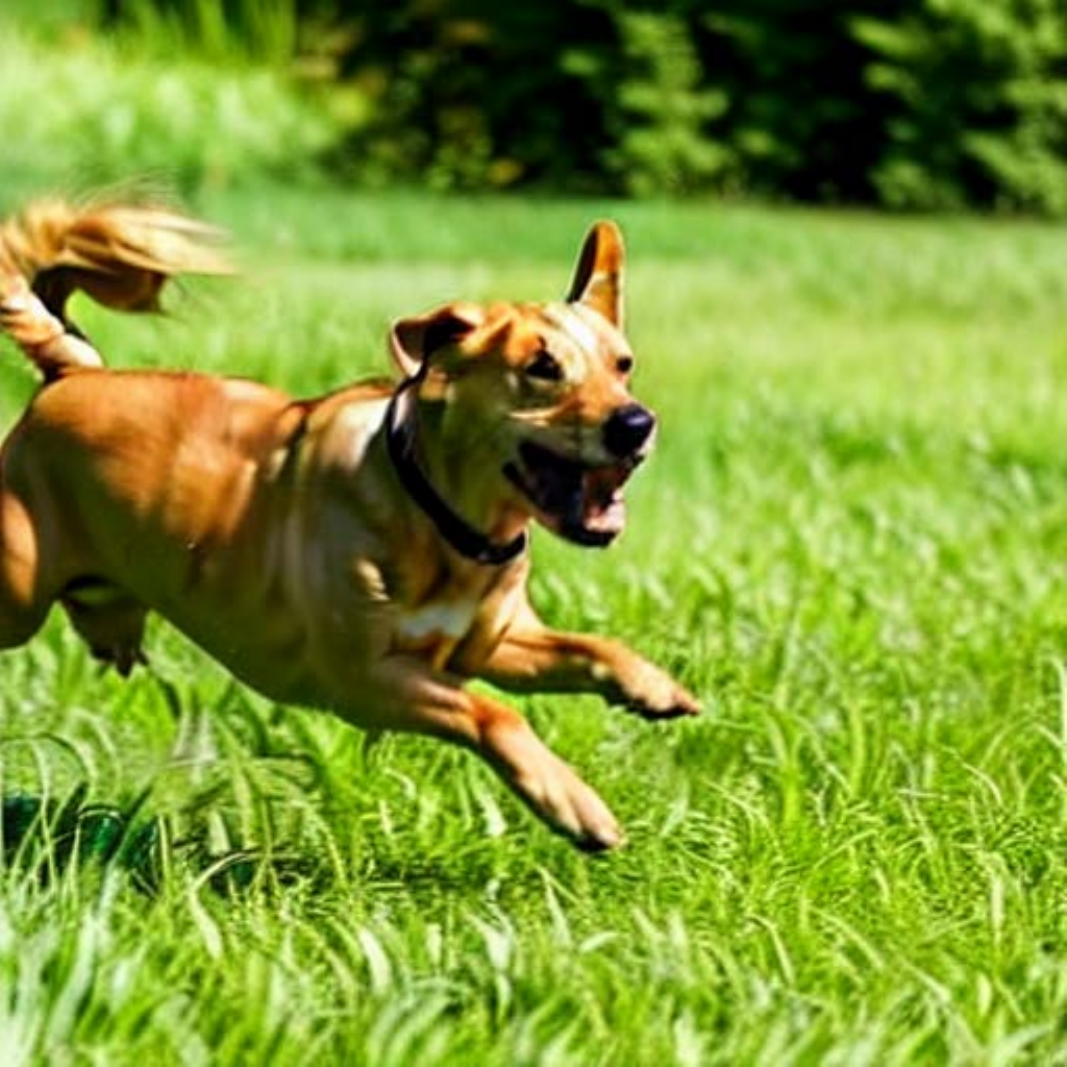} \\

        \includegraphics[width=0.24\linewidth]{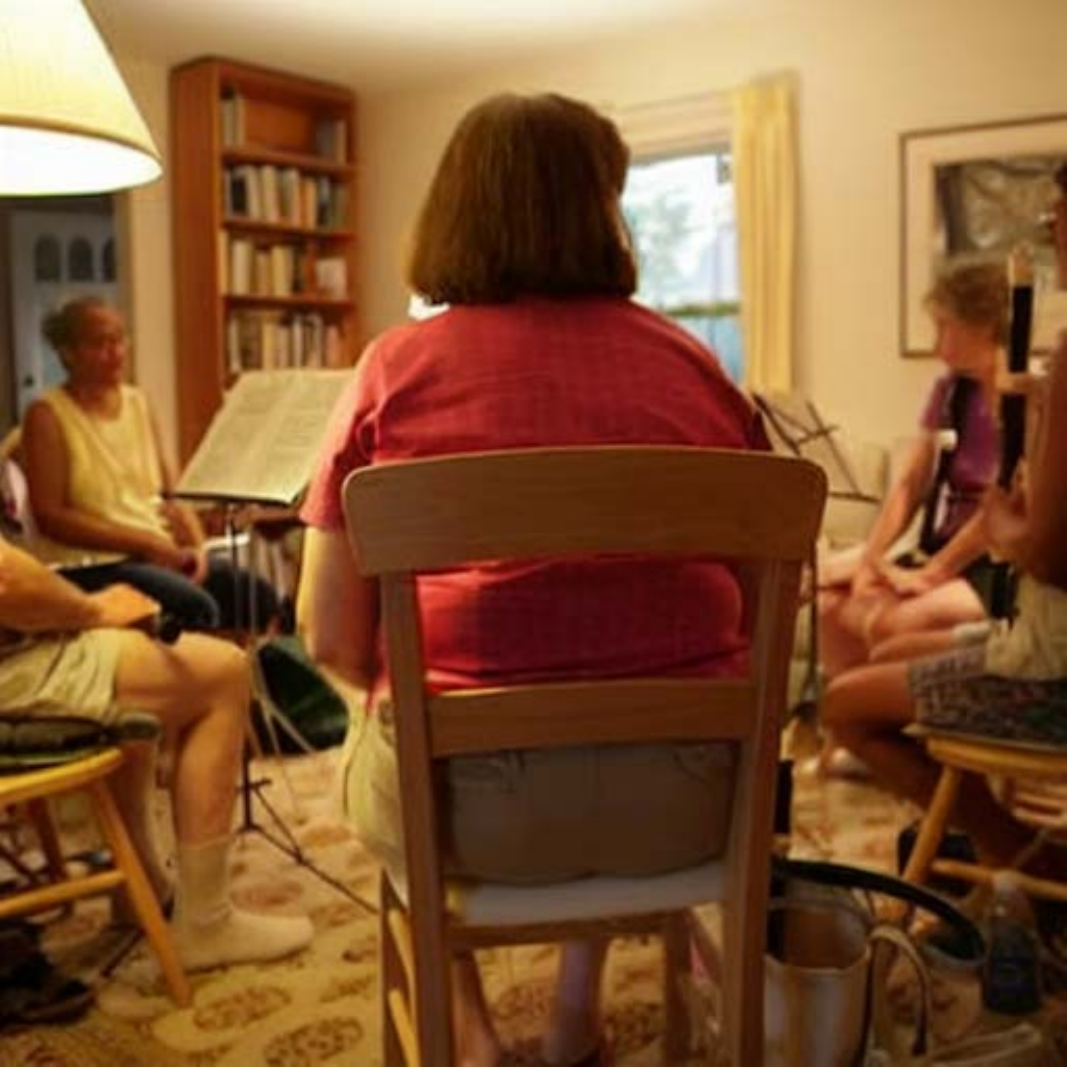} &
        \includegraphics[width=0.24\linewidth]{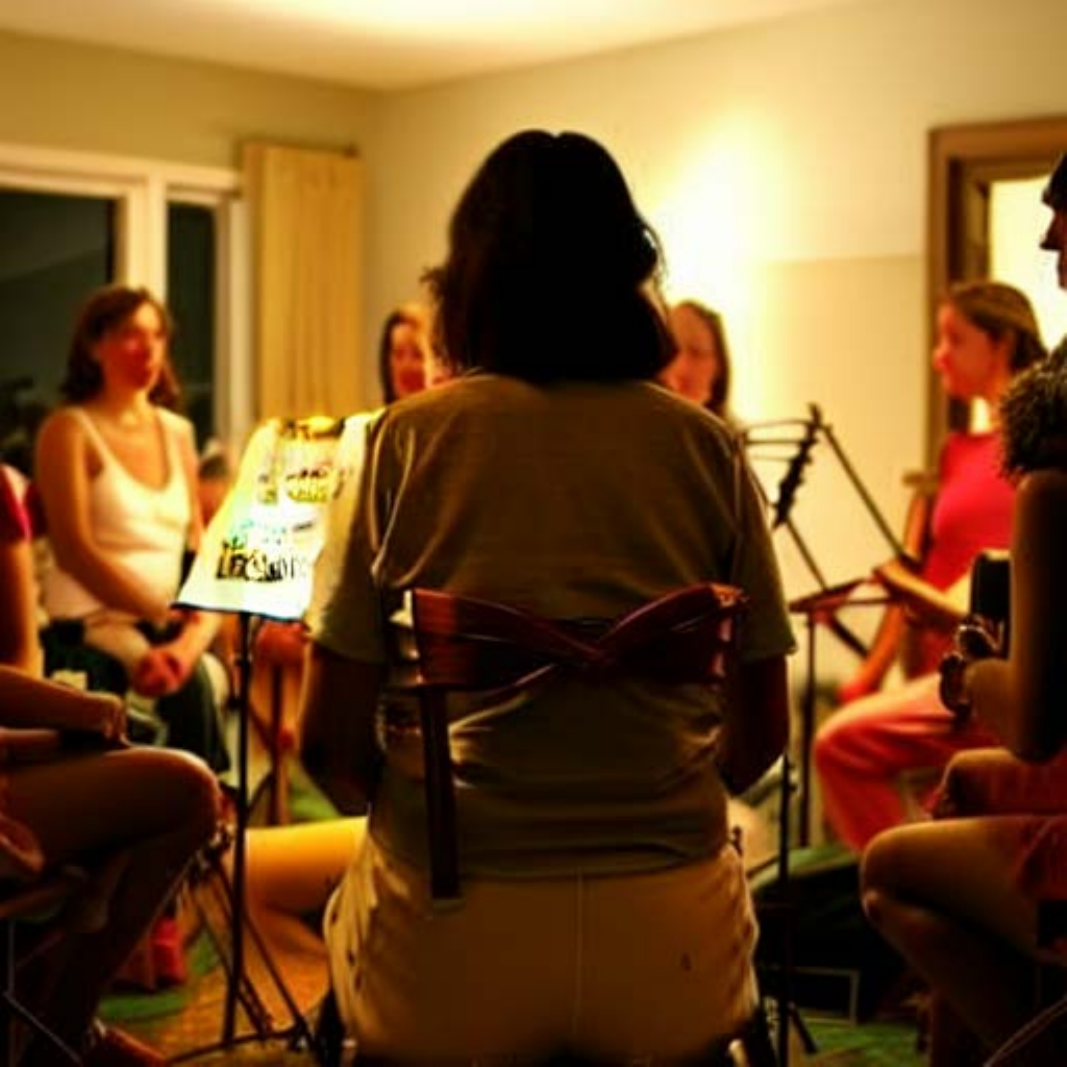} &
        \includegraphics[width=0.24\linewidth]{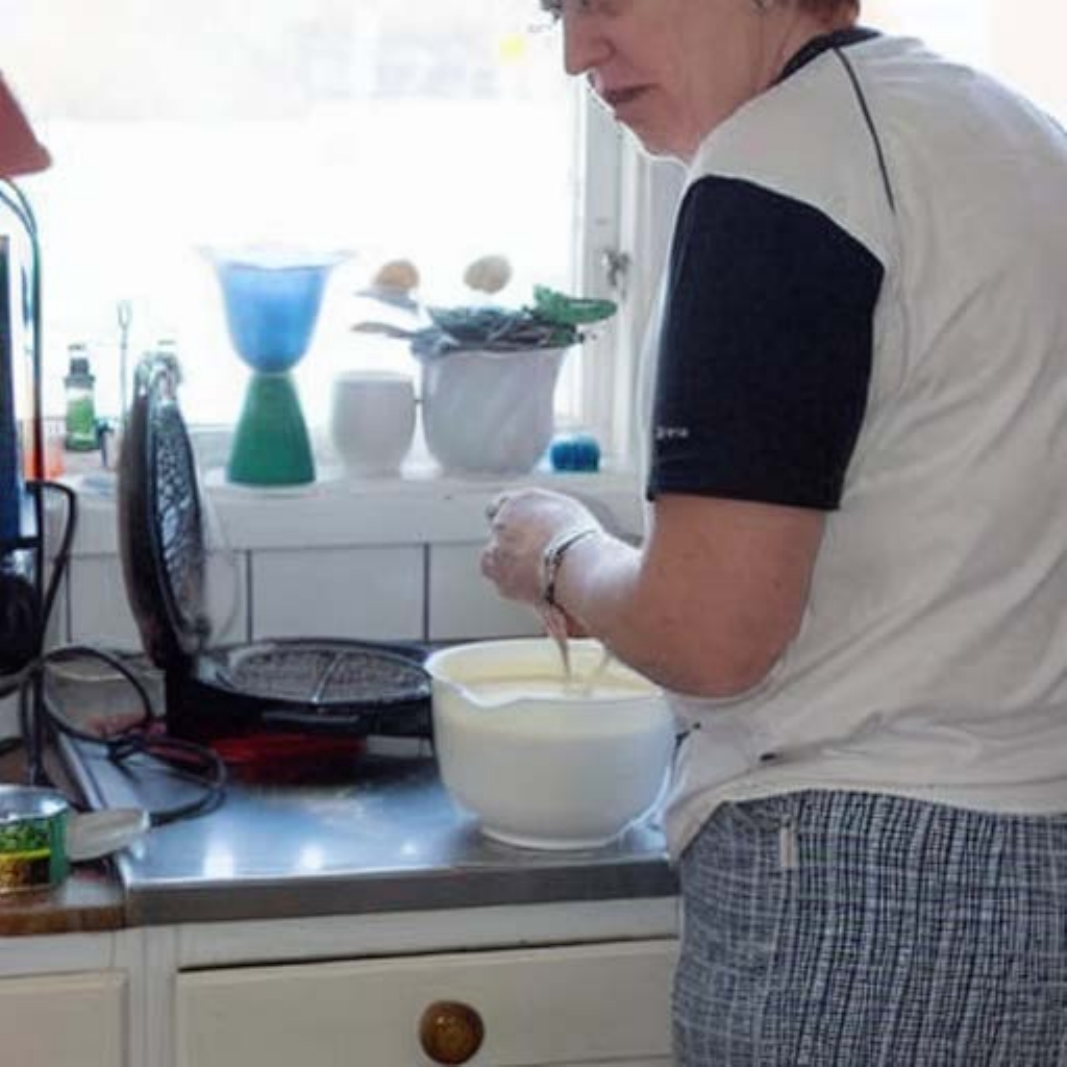} &
        \includegraphics[width=0.24\linewidth]{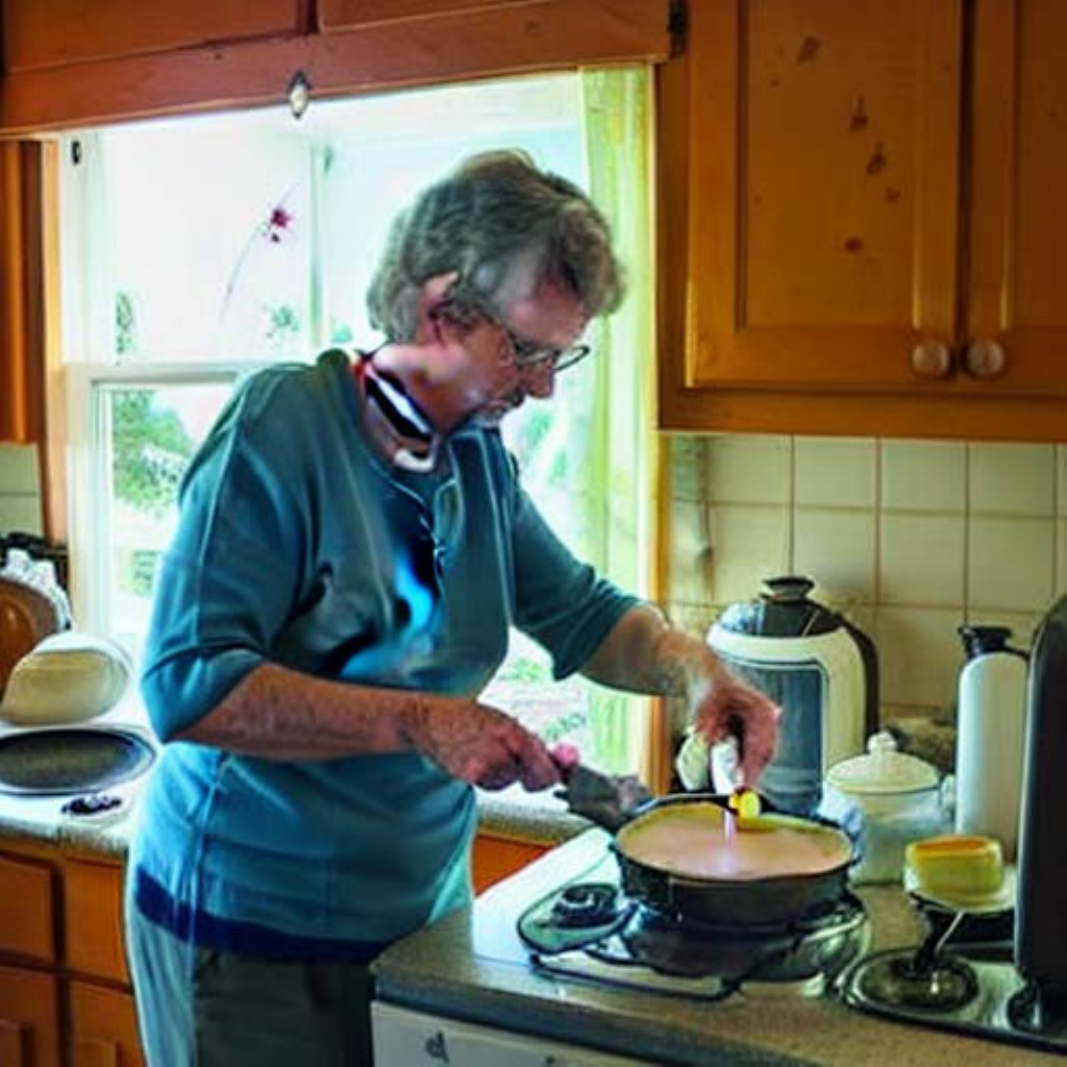} \\

        \includegraphics[width=0.24\linewidth]{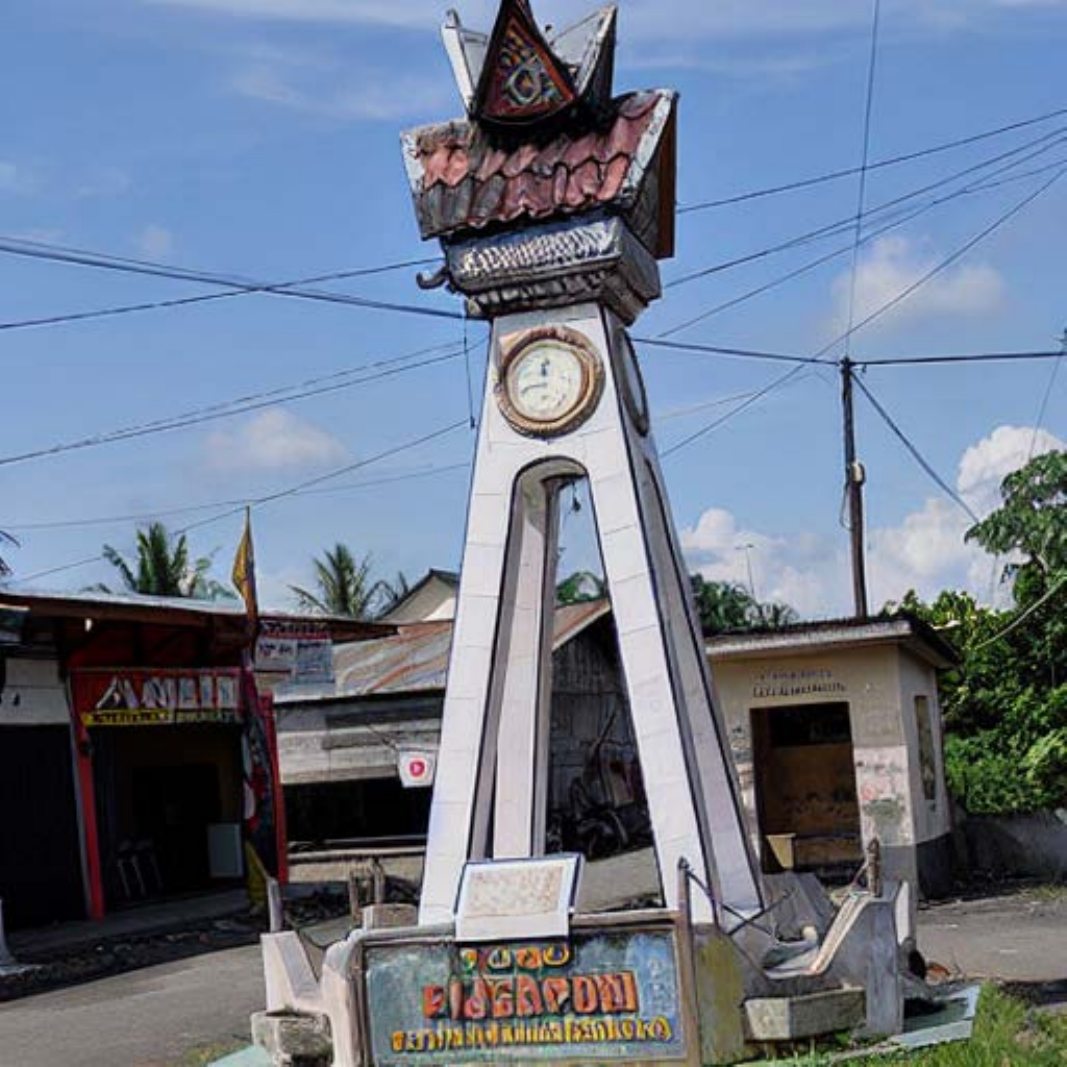} &
        \includegraphics[width=0.24\linewidth]{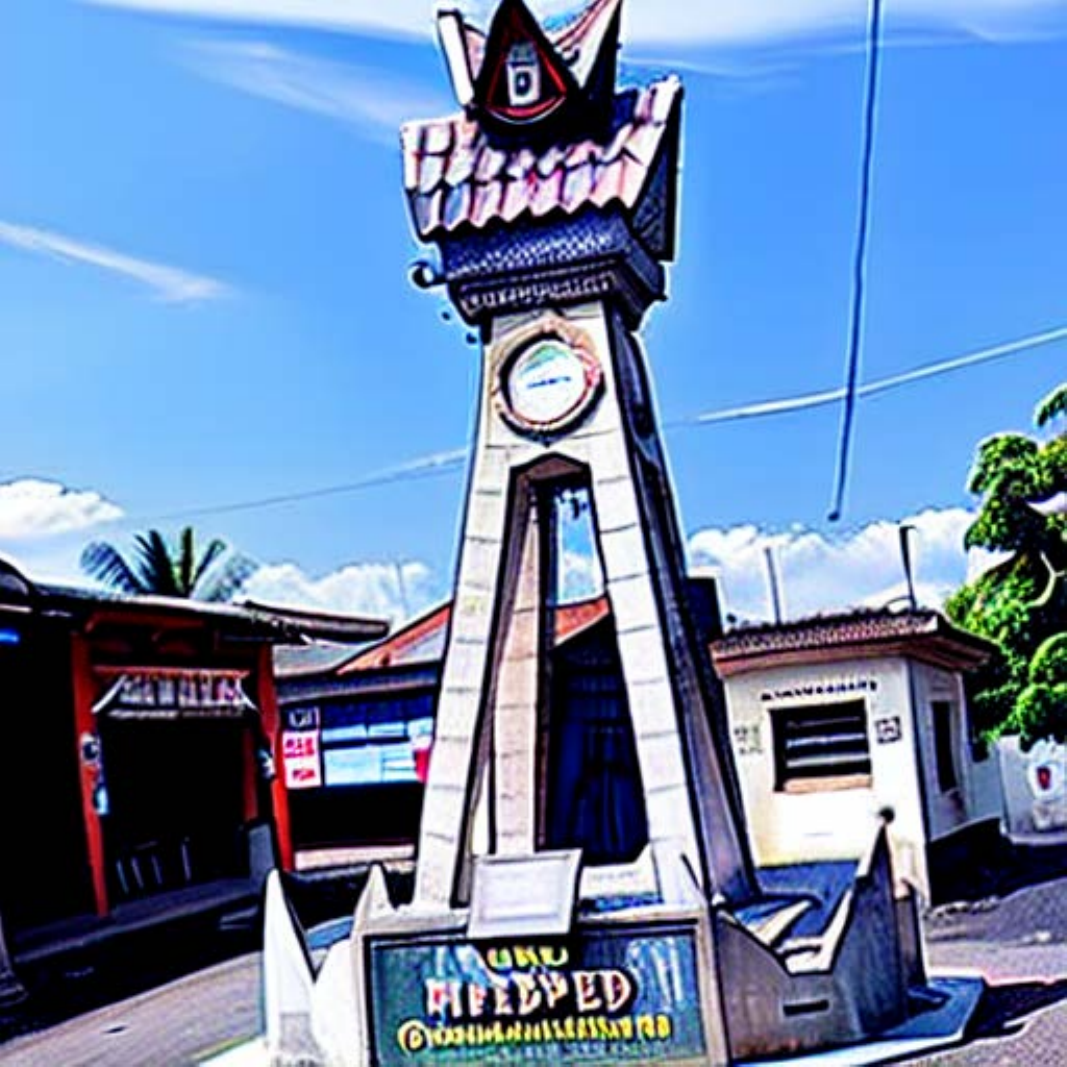} &
        \includegraphics[width=0.24\linewidth]{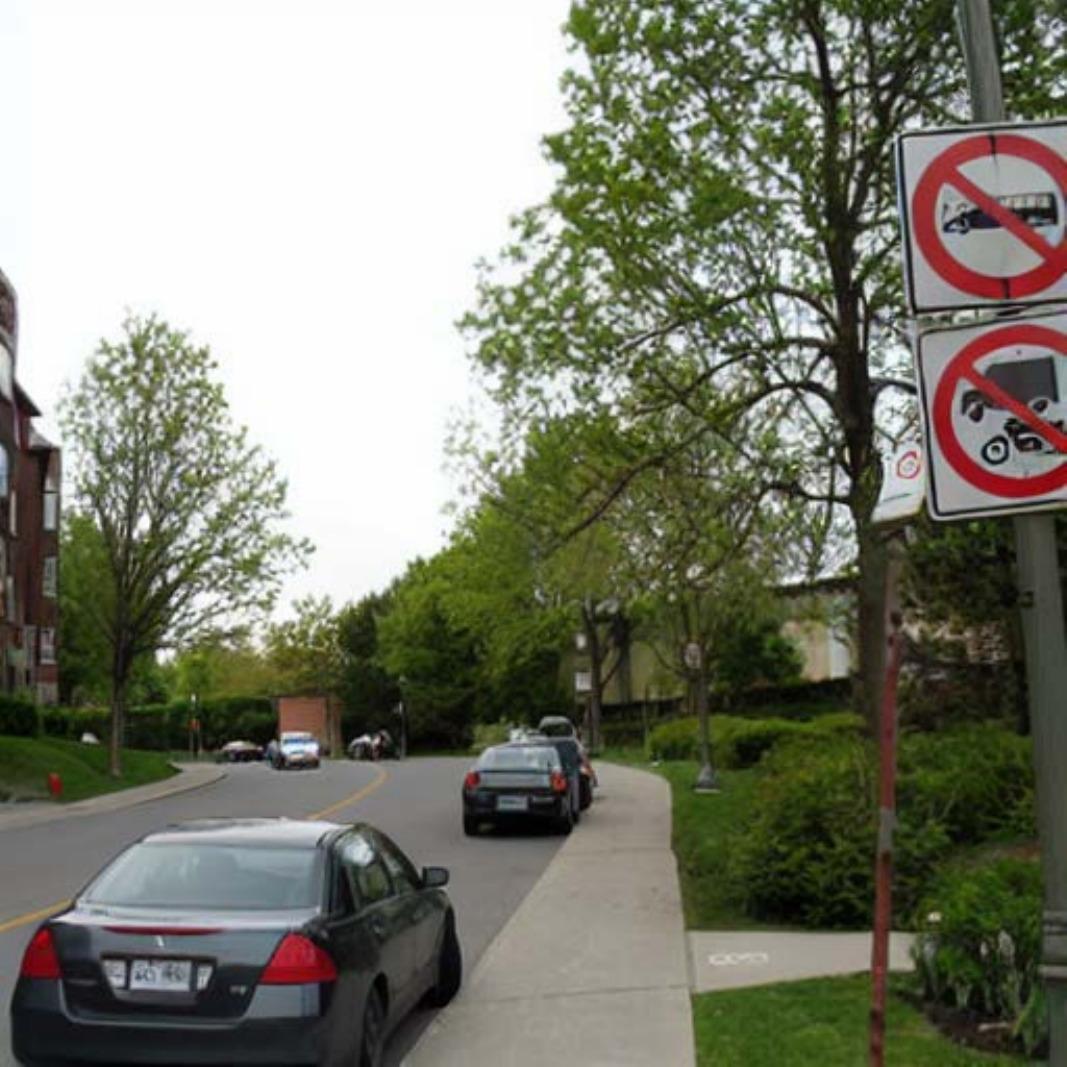} &
        \includegraphics[width=0.24\linewidth]{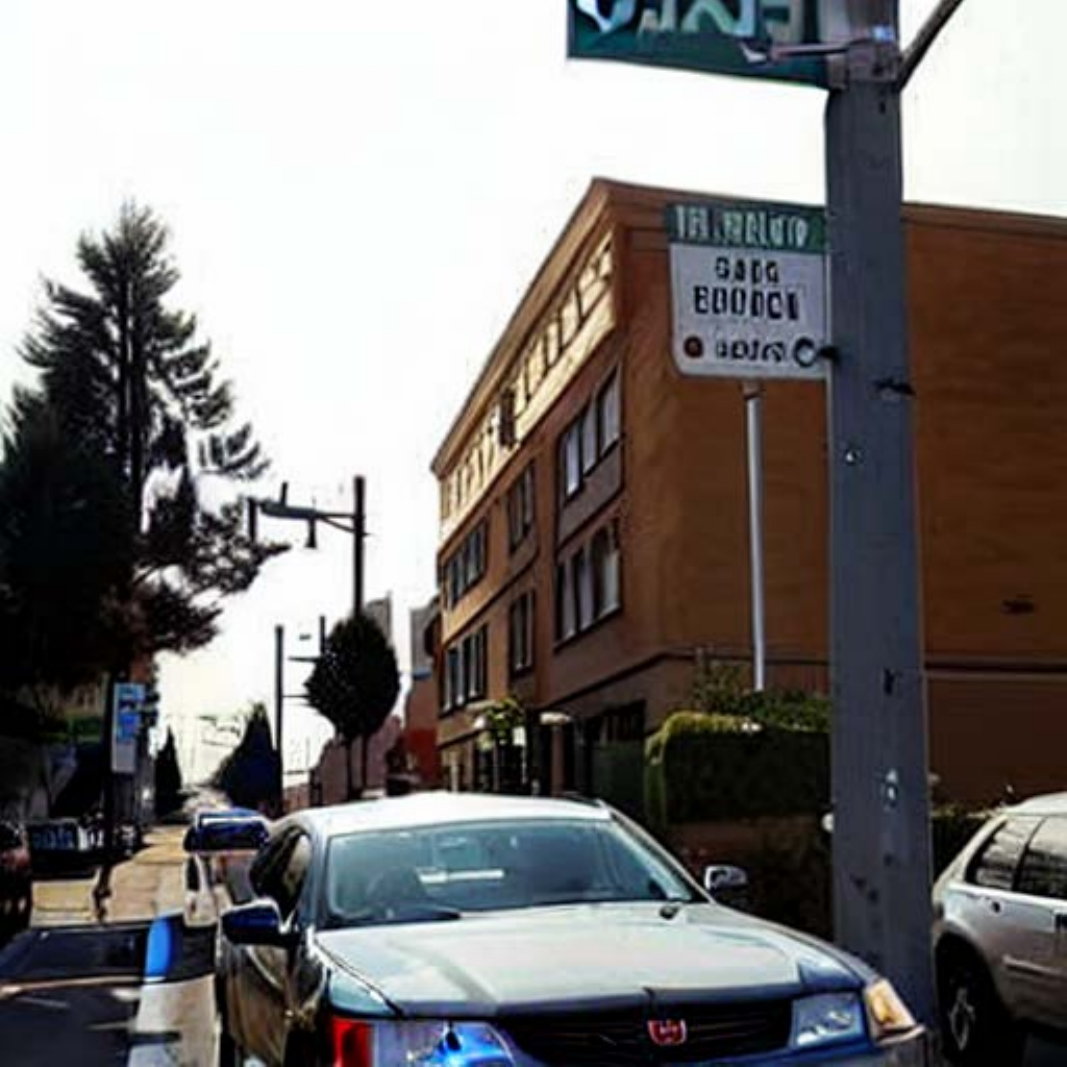} \\
       
    \end{tabular}
    \vspace{-5pt}
    \caption{Visualization of generation results. In our method, the generated images of members are clearly closer to their original images, while the non-members differ significantly from their originals.}
    \label{show}
    \vspace{-7pt}
\end{figure}

Most existing studies \cite{matsumoto2023membership, duan2023diffusion, kong2023efficient, li2024unveiling, zhai2024membership} assumed the adversary can manipulate the intermediate denoising network (denoted as intermediate result attacks). However, this assumption is unrealistic in real-world diffusion systems, which typically expose only end-to-end generation interfaces. Previous end-to-end attacks \cite{pang2023black, wu2022membership} assumed the availability of auxiliary data drawn from the same distribution as the fine-tuning data, which is used to train shadow models and classifiers. However, such attacks depend heavily on the quality of the auxiliary data, and training shadow models and classifiers incurs substantial computational overhead.

We identify a fundamental vulnerability that enables end-to-end attacks without training shadow models or classifiers: standard noise schedules fail to fully eliminate semantic information from images. Analyzing the widely used schedules, we find the signal-to-noise ratio (SNR) at the maximum noise timestep $T$ remains non-zero (Tab.~\ref{SNR}), leaving residual semantic signals \textbf{(Observation 1)}. More critically, via DDIM inversion \cite{dhariwal2021diffusion}, we find that diffusion models inadvertently learn to exploit these residual signals during training, establishing hidden correlations between initial noise and training data, as evidenced by reconstruction fidelity (Tab.~\ref{self_inversion}) and cross-attention analysis (Fig.~\ref{motivation}) \textbf{(Observation 2)}. 

This presents an exploitable attack opportunity: if we can inject an image's semantic information into the initial noise, the model's generation behavior may reveal whether the image belongs to the training set. The key challenge is that adversaries cannot access the target model's denoising network to perform inversion. Fortunately, we observe that fine-tuned models preserve the semantic space of their pre-trained counterparts (Fig.~\ref{motivation} and Tab.~\ref{cosine}), enabling us to use publicly available pre-trained models for semantic injection via DDIM inversion \textbf{(Observation 3)}. 

\begin{figure}[t]
  \centering
  \includegraphics[width=\linewidth]{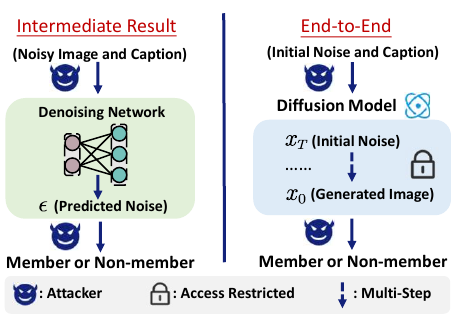}
  \caption{\underline{Left}: the intermediate result attacks, where the adversary supplies inputs to the denoising network and performs attacks based on its predictions. \textit{Note: the diffusion model includes the denoising network, scheduler, and other components.} \underline{Right}: the end-to-end attacks, where the adversary provides inputs to the diffusion model and performs attacks based on the final generation.}
  \label{E-to-E}
\vspace{-5pt}
\end{figure}

Building on these insights, we propose leveraging a pre-trained model to inject semantic information into the initial noise through a DDIM inversion procedure and determine membership by examining the model's generation from semantic initial noise. When the model generates images from this semantic noise, members produce outputs significantly closer to their originals than non-members (Fig.~\ref{show}). This method is an end-to-end attack that requires no access to the target model's parameters or intermediate denoising network. We compare the conditions of our method with those of previous intermediate result attacks, as illustrated in Fig.~\ref{E-to-E}. Furthermore, our method does not need to train shadow models or classifiers. We summarize our contributions as follows:

\begin{itemize}[left=4pt]
    \item To the best of our knowledge, this is the first study to explore the role of initial noise in MIAs against diffusion models. Our key observations indicate that the diffusion model inadvertently captures hidden correlations between the initial noise and the training data, which serves as a crucial indicator for revealing membership information.

    \item Building on this insight, we propose a simple yet effective membership inference attack that leverages the inversion procedure to obtain semantic initial noise. The attack analyzes the model's generation from initial noise containing the original image semantics, with membership determined by the similarity between the generated image and the original image.

    \item Extensive experiments across various datasets validate the effectiveness of our method, with an Area Under the Curve (AUC) of 90.46\% and a True Positive Rate at 1\% False Positive Rate (T@F=1\%) of 21.80\%. It demonstrates that the initial noise can strongly expose membership information, revealing the vulnerability of diffusion models to MIAs.
\end{itemize}

\section{Related Work}
\label{sec:related_work}

\textbf{Membership Inference Attacks.} 
Shokri et al. \cite{MIA} proposed membership inference attacks (MIAs), which primarily targeted classification models in machine learning. The attacks aim to determine whether a specific piece of data has been included in the training set of the target model. MIAs are typically categorized based on the adversary's access level to the target model. In the white-box setting, the attacker is assumed to have full access to the model parameters \cite{leino2020stolen, nasr2019comprehensive, sablayrolles2019white, yeom2018privacy}. In contrast, black-box attacks assume no access to model parameters. Among them, some methods utilize confidence scores or logits provided by the model \cite{MIA, salem2018ml, carlini2022membership}. Other researchers have proposed attacks under a more restrictive assumption where only the final predicted labels are available \cite{choquette2021label, li2021membership, wu2024you}.

\vspace{0.3em} 
\noindent \textbf{Membership Inference Attacks on Diffusion Models.} 
The inherent generative mechanism of diffusion models substantially increases the difficulty of membership inference and renders previous approaches largely ineffective \cite{duan2023diffusion}. Recently, MIAs on diffusion models have garnered increasing attention. Pang et al. \cite{pang2023white} proposed executing an attack by utilizing gradient information. Recent works \cite{matsumoto2023membership, duan2023diffusion, kong2023efficient, li2024unveiling, zhai2024membership, lian2025unveiling} assumed that the adversary has access to the model's intermediate denoising process and is allowed to modify the inputs of the denoising networks. This assumption enables queries on the denoising networks' prediction to infer membership information. Some works \cite{pang2023black, wu2022membership} relied on an auxiliary dataset drawn from the same distribution to train shadow models, and trained a classifier based on the behavioral differences of the shadow models for member and non-member samples. Zhang et al. \cite{zhang2024generated} assumed access to non-member data and constructed member samples through extensive queries for the attack.

\vspace{0.3em} 
\noindent \textbf{Denoising Diffusion Implicit Model (DDIM) Inversion.} DDIM Inversion is a technique that utilizes the reverse process of the diffusion model to obtain the initial state of the generated image \cite{dhariwal2021diffusion}. In addition, it can be viewed as a noise addition process that integrates semantic information \cite{lichenzigzag, zhou2024golden}. It adds the predicted noise to the original image, resulting in an initial noise that retains the semantics of the original image. In contrast, naively adding random noise to the image is not connected to the model's understanding of the original image's semantics \cite{zhang2023inversion}. Due to the unique advantages of DDIM inversion, it shows great potential in applications such as image quality optimization \cite{lichenzigzag, zhou2024golden} and image editing \cite{zhang2023inversion, garibi2024renoise, dong2023prompt}.

\section{Threat Model}
\label{threat}
MIAs aim to determine whether a specific sample was used during the model training. Formally, let \(G_\theta\) be a fine-tuned diffusion model with parameters \(\theta\). Let \(D\) be a dataset drawn from the data distribution \(q_{\mathrm{data}}\) and each sample \(x_i\) in \(D\) has a caption $\mathbf{c}_i$. Following established conventions \cite{sablayrolles2019white, carlini2022membership, duan2023diffusion}, we split \(D\) into two subsets \(D_M\) and \(D_N\), where \(D_M\) denotes the member set used to fine-tune the diffusion model \(G_\theta\) and \(D_N\) denotes the non-member set, so that \(D = D_M \cup D_N\) and \(D_M \cap D_N = \varnothing\). Each image sample \(x_i\) is associated with a membership label \(m_i\), where \(m_i=1\) if \(x_i\in D_M\) and \(m_i=0\) otherwise. The adversary has access to the dataset \(D\), but does not know the partition.  

\vspace{0.3em} 
\noindent \textbf{Adversary's Goal.} The adversary's goal is to design a membership inference attack algorithm \(\mathcal{A}\) that, for any given sample \(x_i\), predicts its corresponding membership label:
\begin{equation}
\mathcal{A}(x_i, \theta) = \mathds{1} \left[ \mathbb{P}(m_i = 1 \mid \theta, x_i) \geq \tau \right],
\end{equation}
where \( \mathcal{A}(x_i, \theta) = 1 \) means \( x_i \) comes from \( D_M \), \(  \mathds{1} [A] = 1 \) if \( A \) is true, and \( \tau \) is the threshold. 

\vspace{0.3em} 
\noindent \textbf{Adversary's Capabilities.} In this paper, the adversary is limited to performing end-to-end generation using the diffusion model. Specifically, the adversary can modify the model's initial noise and prompt, and can only observe the final generated image, but has no access to any model parameters or intermediate denoising steps. In the diffusers library \cite{von-platen-etal-2022-diffusers}, pipeline interfaces can accept an initial noise input, which can be adjusted to guide the generation process. Moreover, tasks such as image editing \cite{zhou2024golden, wang2024silent, sun2024spatial}, and noise engineering \cite{guo2024initno, mao2023guided, garibi2024renoise} also rely on interfaces that allow the modification of the initial noise. The fine-tuned model is obtained by fine-tuning a pre-trained model on a downstream dataset. Following prior membership inference settings for fine-tuned models \cite{fu2024membership, pang2023black, zhai2024membership, duan2023diffusion}, we assume that the adversary has access to the pre-trained version of the fine-tuned model.\footnote{In the following sections, we demonstrate that our method remains effective even when the pre-trained version is unknown.} We compare the capabilities of adversaries across different attack algorithms in Tab.~\ref{Adversary}. 


\begin{table}[h]
\centering
\vspace{-5pt}
\caption{Adversary capabilities for different attacks. The top half does not apply to end-to-end generation; the bottom half is feasible. \underline{Pm}: access to model parameters; \underline{Inter}: control the inputs to the denoising network at intermediate timesteps; \underline{Shadow}: train shadow models; \underline{Arch}: known fine-tuning architecture version; \underline{cls}: train classifiers. \underline{Aux}: access to auxiliary dataset. \textit{Symbols}: \cmark = required, \xmark = not required.} 
\label{Adversary}
\resizebox{0.47\textwidth}{!}{  
\begin{tabular}{c c c c c c c}
\toprule
\textbf{Method} & \textbf{Pm} & \textbf{Inter} & \textbf{Shadow} & \textbf{Arch} & \textbf{Cls} & \textbf{Aux} \\
\midrule
GSA \cite{pang2023white}  & \cmark & \cmark & \cmark & \cmark & \cmark & \cmark\\
Loss \cite{matsumoto2023membership} & \xmark & \cmark & \xmark & \cmark & \xmark & \cmark\\
SecMI \cite{duan2023diffusion} & \xmark & \cmark & \xmark & \cmark & \xmark & \cmark\\
PIA \cite{kong2023efficient} & \xmark & \cmark & \xmark & \cmark & \xmark & \cmark\\
CLiD \cite{zhai2024membership} & \xmark & \cmark & \cmark & \cmark & \xmark & \cmark\\
\midrule  
NA-P \cite{wu2022membership} & \xmark & \xmark & \cmark & \cmark & \cmark & \cmark\\
Feature-T \cite{pang2023black} & \xmark & \xmark & \cmark & \xmark & \xmark & \cmark\\
Feature-C \cite{pang2023black} & \xmark & \xmark & \cmark & \cmark & \cmark & \cmark\\
Feature-D \cite{pang2023black} & \xmark & \xmark & \cmark & \cmark & \cmark & \cmark\\
\textcolor{red}{\textbf{Ours}} & \textcolor{red}{\xmark} & \textcolor{red}{\xmark} & \textcolor{red}{\xmark} & \textcolor{red}{\cmark} & \textcolor{red}{\xmark} & \textcolor{red}{\cmark} \\
\bottomrule
\end{tabular}
}
\vspace{-7pt}
\end{table}

\section{Methodology}
\begin{figure*}[h]
	\centering	
	\includegraphics[width=6.80in]{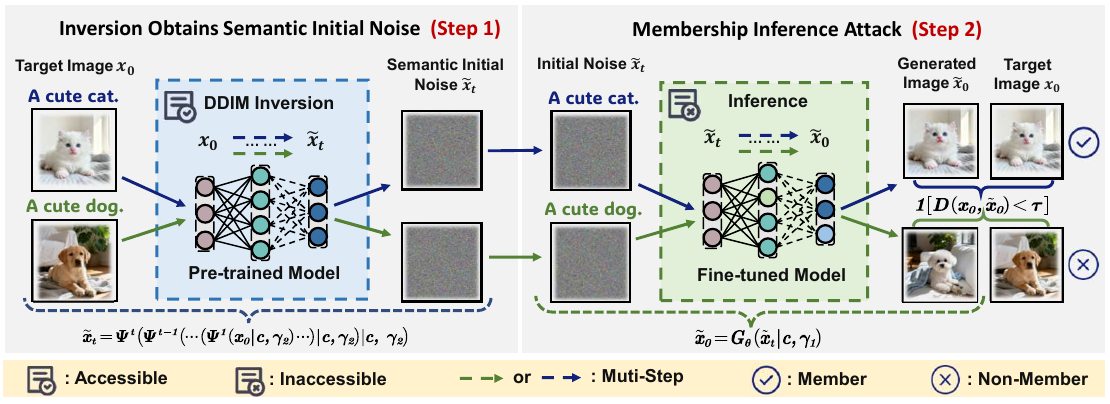}	
    \caption{Overview of our method. \underline{Step 1}: Use a pre-trained model for DDIM inversion to obtain initial noise with semantics. \underline{Step 2}: Generate images using the noise and determine membership based on the generation results.}
	\label{fig_idea}
    \vspace{-3pt}
\end{figure*}

In this section, we present our three key observations systematically. Based on these observations, we propose a membership inference attack that leverages initial noise, which is performed in two steps, as shown in Fig.~\ref{fig_idea}.

\subsection{Diffusion Models}
Given an original image \(x_0\), the forward diffusion process gradually introduces noise over \(T\) timesteps, transforming \(x_0\) into a nearly Gaussian-distributed \(x_T\). The forward diffusion process at each timestep \(t\) is described as follows:
\begin{equation}
x_t = \sqrt{\bar{\alpha}_t} x_0 + \sqrt{1 - \bar{\alpha}_t} \epsilon,
\end{equation}
where \( \overline{\alpha}_t = \prod_{i=1}^{t} \alpha_i \) and \( (\alpha_1, \ldots, \alpha_T) \) are the predefined noise schedules, \(\epsilon \sim \mathcal{N}(0, I)\) represents Gaussian noise at each step. As \(t\) approaches \(T\), \(\bar{\alpha}_t\) gradually diminishes, making \(x_T \approx \epsilon\) closely resemble pure Gaussian noise. In this paper, we use $\epsilon_\theta$ to denote the prediction of the denoising network.

\subsection{The Semantics in Initial Noise}
\colorbox{gray!20}{\parbox{0.465\textwidth}{\textbf{Observation 1.} The widely adopted noise schedules fail to eliminate the semantic information in the original image, even at the maximum noise step.}}

\vspace{0.3em} 
\noindent \textbf{Explanation 1.} We begin with the forward noise addition process in diffusion models. We follow the definition of the signal-to-noise ratio (SNR) in the diffusion model training process from prior work \cite{choi2022perception}. For a given timestep $t$, SNR can be characterized as follows: 
\begin{equation}
\text{SNR}(t) := \bar{\alpha}_t/(1 - \bar{\alpha}_t).
\end{equation}
We compared the SNR of images at step \(T\) across different schedules in Tab.~\ref{SNR}. Consistent with the analyses in prior work \cite{wang2024silent, lin2024common}, noise injection at step \(T\) cannot eliminate the original signal. 

\begin{table}[h]
\caption{Comparison of different noise schedules in the final signal-to-noise ratio $\text{SNR}(T)$ and the corresponding $\sqrt{\bar{\alpha}_T}$. The results show that, despite large differences across schedules (Linear \cite{DDIM}, Cosine \cite{nichol2021improved}, and Stable Diffusion \cite{rombach2022high}), residual signals consistently remain at the last step.}
\label{SNR}
\centering
\begin{tabular}{l|cc}
\hline
Schedule   & $\text{SNR}(T)$   & $\sqrt{\bar{\alpha}_T}$ \\ \hline
Linear   & 4.04e-05      & 0.006353                \\
Cosine   & 4.24e-09     & 0.00004928         \\
Stable Diffusion & 4.68e-03          & 0.068265                \\ \hline
\end{tabular}
\vspace{-3pt}
\end{table}

The diffusion model's training process can be interpreted as learning a transformation from a Gaussian noise distribution to an image distribution. Previous studies on noise engineering \cite{zhou2024golden, wang2024silent, lichenzigzag, sun2024spatial} have shown that the initial noise contains semantic information that influences the generation process. Based on the evidence above, we speculate that the semantics of the initial noise may be linked to the training data, potentially revealing membership information. Our subsequent observation further supports this speculation.

\vspace{0.3em} 
\noindent \colorbox{gray!20}{\parbox{0.465\textwidth}{\textbf{Observation 2.} The diffusion model inadvertently learns to exploit residual information in the initial noise, thereby establishing a hidden connection between the initial noise and the training data during training.}}

\vspace{0.3em} 
\noindent \textbf{Explanation 2(a).} As discussed in Sec.~\ref{sec:related_work}, DDIM inversion can be regarded as a noise injection process that embeds the semantics of the original image into the initial noise. The inversion process \(\Psi\) can be expressed as follows:
\begin{equation}
\begin{split}
\label{inversion}
\tilde{x}_t = \Psi^t(\tilde{x}_{t-1} | \mathbf{c}, \gamma_2) 
= \sqrt{\overline{\alpha}_{t}} &\tilde{f}_\theta(\tilde{x}_{t-1}, t-1) \\
&+ \sqrt{1 - \overline{\alpha}_{t}}\epsilon_\theta(\tilde{x}_{t-1}, t-1),
\end{split}
\end{equation}
where $\tilde{f}_\theta(\tilde{x}_{t-1}, t-1)$ can be expressed as:
\begin{equation}
\tilde{f}_\theta(\tilde{x}_{t-1}, t-1) = \frac{\tilde{x}_{t-1}-\sqrt{1 - \overline{\alpha}_{t-1}} \epsilon_\theta(\tilde{x}_{t-1}, t-1)}{\sqrt{\overline{\alpha}_{t-1}}},
\end{equation}
where \(\epsilon_\theta(\tilde{x}_{t-1}, t-1) = (1+\gamma_2) \epsilon_\theta(\tilde{x}_{t-1}, \mathbf{c}, t-1) - \gamma_2 \epsilon_\theta(\tilde{x}_{t-1}, \varnothing, t-1)\), $\mathbf{c}$ is the text prompt, $\varnothing$ represents the null prompt and \(\gamma_2\) is the inversion guidance scale. 
Analyzing the diffusion model's training process and the characteristics of DDIM inversion, we attempt to use the target model for DDIM inversion to obtain semantic initial noise, which can be expressed as:
\begin{equation}
\small
\begin{split}
\tilde{x}_t & = Inv^t_{\theta}(x_0|\mathbf{c}, \gamma_2) \\ 
&= \Psi^t\Big( \Psi^{t-1}\big( \cdots (\Psi^1(x_0 \mid \mathbf{c}, \gamma_2)\cdots)\mid \mathbf{c},\gamma_2\big)\mid \mathbf{c},\gamma_2 \Big).
\end{split}
\end{equation} 
Then, we use the obtained noise as the starting point for image generation, which can be formulated as:
\begin{equation}
\tilde{x}_0=G_\theta(\tilde{x}_t|\mathbf{c}, \gamma_1).
\end{equation}
where $\gamma_1$ is the guidance scale during generation. We compared the normalized $\ell_{2}$ distance between the original images and those generated from random noise or semantic noise (obtained via inversion). Tab.~\ref{self_inversion} reports the statistical results for both member and non-member samples across different datasets. The results demonstrate that semantic initial noise leads to higher fidelity reconstructions of the training data. This confirms that diffusion models capture residual information from images, and when conditioned on initial noise containing member semantics, they generate outputs that are closer to the original member images.

\begin{table}[h]
    \centering
    \caption{Statistics of normalized $\ell_{2}$ distance across different datasets. $Random$: generate from random noise, $Inversion$: generate from semantic noise, $\Delta=Inversion - Random$.}

    \label{self_inversion}
    \renewcommand{\arraystretch}{1.0} 
    \resizebox{0.48\textwidth}{!}{
    \begin{tabular}{@{}lccccccc@{}}
        \toprule
        \multirow{2}{*}{\makecell{Method}} & \multicolumn{2}{c}{Pokémon} & \multicolumn{2}{c}{MS-COCO} & \multicolumn{2}{c}{Flickr} \\ 
        \cmidrule(lr){2-3} \cmidrule(lr){4-5} \cmidrule(lr){6-7}
        & Mem & Non-Mem & Mem & Non-Mem & Mem & Non-Mem \\ 
        \midrule
        $Random$ & 0.3839 & 0.3961 & 0.4034 & 0.4963 & 0.3543 & 0.3972  \\ 
        $Inversion$ & 0.2779 & 0.5061 & 0.2469 & 0.5394 & 0.2722 & 0.4464 \\
        \midrule
        $\Delta$ & -0.1060 & +0.1100 & -0.1565 & +0.0431 & -0.0821 & +0.0492 \\
        \bottomrule
    \end{tabular}
    }

\end{table}

\begin{figure}
    \centering
    \includegraphics[width=0.48\textwidth]{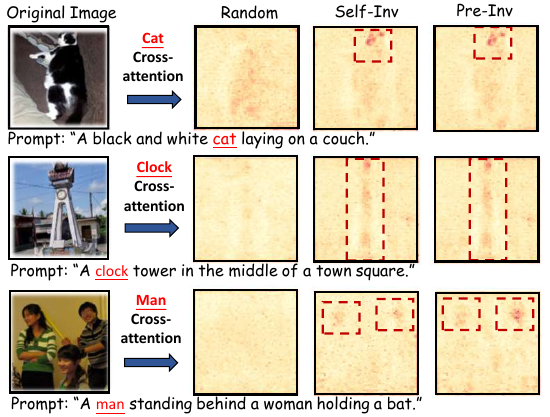}
    \caption{Visualization of cross-attention heatmaps. Heatmaps display the local contributions of the second attention modules in the third upsampling block. \textbf{Random}: generation using random noise; \textbf{Self-Inv}: generation using semantic noise obtained via inversion of the target model; \textbf{Pre-Inv}: generation using semantic noise obtained via inversion of the pre-trained model. The red boxes highlight regions with high attention, which precisely correspond to the locations of the main objects in the original images.}
    \label{motivation}
    \vspace{-5pt}
\end{figure}

\vspace{0.3em} 
\noindent \textbf{Explanation 2(b).} Previously, we observed that DDIM inversion enables better reconstruction of member data. We attribute this to the model having learned correlations between residual semantics and the original images during training. To further validate this hypothesis, we analyze the cross-attention of the denoising network during the generation process. Specifically, we visualize the cross-attention heatmaps at the first denoising step, which compute the attention of the token corresponding to the main object in the image. As shown in Fig.~\ref{motivation}, when using initial noise obtained via DDIM inversion, the denoising network immediately attends to the semantic regions of the image. In contrast, when initialized with random noise, the network exhibits no clear focus on specific regions. This phenomenon further supports that the model has inadvertently established a hidden connection between the initial noise and the original image. 

Although inversion provides semantic initial noise for both member and non-member samples, members can be reconstructed much more faithfully because the model has learned correlations between such semantic cues and the original training images. As a result, the semantic information contained in the initial noise allows the model to better recover the original images of members, making it a discriminative signal for membership inference.

\subsection{Semantic Injection via Pre-trained Model}
One major challenge is that the adversary cannot directly access the parameters of the target model, making it infeasible to perform DDIM inversion on the target itself. Despite these constraints, we observe a crucial aspect of fine-tuning in the context of membership inference attacks.

\vspace{0.3em} 
\noindent \colorbox{gray!20}{\parbox{0.465\textwidth}{\textbf{Observation 3.} Models fine-tuned from pre-trained ones essentially preserve the original semantic space and representational capabilities.}}

\vspace{0.3em} 
\noindent \textbf{Explanation 3.} Previous works \cite{zhou2022closer, radiya2020fine} suggested that fine-tuning does not substantially alter the original distribution. Based on this characteristic, we hypothesize that the model's understanding and representation of semantic information have not undergone significant changes after fine-tuning. To verify this hypothesis, we perform inversion with the pre-trained version of the target model to obtain semantic initial noise. Following the same procedure as before, we then feed this noise into the target model and examine its cross-attention heatmaps. As shown in Fig.~\ref{motivation}, the fine-tuned model still captures the semantics in the initial noise, with high cross-attention similarity to that obtained from self-inversion\footnote{For clarity, DDIM inversion with the target model is termed self-inversion, and that with the pre-trained model is termed pre-inversion.}. 

\begin{table}[h]
    \centering
    \caption{Similarity of cross-attention heatmaps. Across different datasets and fine-tuning epochs, the cross-attention heatmaps show a remarkably high similarity between the noises obtained from self-inversion and pre-inversion.}

    \label{cosine}
    \renewcommand{\arraystretch}{1.0} 
    \resizebox{0.48\textwidth}{!}{
    \begin{tabular}{@{}lcccccc@{}}
        \toprule
        Epoch & 50 & 100 & 150 & 200 & 250 & 300 \\ 
        \midrule
        Pokémon & 0.968 & 0.942 & 0.926 & 0.897 & 0.895 & 0.892 \\ 
        T-to-I & 0.970 & 0.954 & 0.924 & 0.899 & 0.893 & 0.893  \\
        MS-COCO & 0.973 & 0.953 & 0.937 & 0.897 & 0.895 & 0.895  \\
        Flickr & 0.978 & 0.953 & 0.934 & 0.904 & 0.902 & 0.903 \\
        \bottomrule
    \end{tabular}
    }

\end{table} 

To further confirm this similarity, we computed the cosine similarity of cross-attention heatmaps between the initial noise obtained from self-inversion and pre-inversion. As shown in Tab.~\ref{cosine}, across different fine-tuning epochs, the cross-attention heatmaps of the noise from the two inversions exhibit a high degree of similarity. All the findings provide evidence that the fine-tuned model retains the semantic space of its pre-trained version, making it possible to obtain semantic initial noise using a pre-trained model.

\begin{table*}[h]
    \centering
    \caption{AUC and T@F=1\% on different datasets. The compared baselines are divided into two categories: intermediate result attacks (\textit{Upper part}) and end-to-end attacks (\textit{Lower part}). In each column, the best performance in the end-to-end attacks is displayed in \underline{bold}, while the best performance across all attacks is \underline{underlined}. Our method achieves the best performance in end-to-end attacks and demonstrates performance comparable to that of intermediate result attacks. Notably, unlike intermediate result attacks, our method does not require access to the denoising network's inputs or outputs.}

    \label{Performance}
    \renewcommand{\arraystretch}{1.0} 
    \begin{tabular}{@{}lccccccccccc@{}}
        \toprule
        \multirow{2}{*}{\makecell{Method}} & \multicolumn{2}{c}{Pokémon} & \multicolumn{2}{c}{T-to-I} & \multicolumn{2}{c}{MS-COCO} & \multicolumn{2}{c}{Flickr} & \multicolumn{2}{c}{Average} \\ 
        \cmidrule(lr){2-3} \cmidrule(lr){4-5} \cmidrule(lr){6-7} \cmidrule(lr){8-9} \cmidrule(lr){10-11}
        & AUC & T@F=1\% & AUC & T@F=1\% & AUC & T@F=1\% & AUC & T@F=1\% & AUC & T@F=1\% \\ 
        \midrule
        SecMI & \underline{83.26} & 12.88 & 88.26 & \underline{26.66} & 89.37 & 16.79 & \underline{76.42} & 14.40 & 84.33& 17.68\\ 
        PIA & 76.82 & 7.85 & 84.80 & 11.78 & 71.38 & 5.20 & 72.59 & 7.20 & 76.40 & 8.01 \\
        \midrule
        NA-P & 59.37 & 4.80 & 70.77 & 6.60 & 52.41 & 2.20 & 56.51 & 4.00 & 59.77 & 4.40 \\ 
        GD & 52.67 & 1.20 & 59.67 & 5.20 & 51.22 & 1.00 & 53.41 & 2.00 & 54.24 & 2.35 \\ 
        Feature-T & 56.60 & 3.00 & 70.40 & 7.00 & 57.70 & 3.20 & 58.20 & 3.00 & 60.73 & 4.05 \\      
        Feature-C & 60.67 & 7.33 & 83.77 & 17.40 & 73.08 & 14.60 & 63.80 & 5.20 & 70.33 & 11.13 \\ 
        Feature-D & 55.10 & 2.80 & 65.00 & 6.00 & 58.00 & 3.00 & 57.20 & 3.00 & 58.83 & 3.70 \\
        \rowcolor[gray]{0.9}
        \textbf{Ours} & \textbf{82.44} & \underline{\textbf{14.00}} & \underline{\textbf{89.24}} & \textbf{21.60} & \underline{\textbf{90.46}} & \underline{\textbf{21.80}} & \textbf{76.23} & \underline{\textbf{16.00}} & \underline{\textbf{84.59}} & \underline{\textbf{18.35}} \\     
        \bottomrule
    \end{tabular}
    \vspace{-3pt}
\end{table*}

\subsection{MIAs Leveraging Initial Noise}
Based on the above experiments and observations, we propose a simple yet effective membership inference attack that exploits the correlations between the semantic initial noise and the training data. This attack proceeds in two main steps: obtaining semantic initial noise and conducting the membership inference attack. \textit{We provide the detailed algorithmic procedure in algorithm~\ref{algorithm_idea}.}

\begin{algorithm}[h]
\caption{MIAs Leveraging Initial Noise}
\label{algorithm_idea}
\raggedright\textbf{Input:} Target model $G_{\theta}$, pre-trained model $G_{\theta_{\text{pre-trained}}}$, threshold $\tau$, target data $(x, \mathbf{c})$, distance metric $D(\cdot, \cdot)$, inference guidance scale $\gamma_1$, inversion guidance scale $\gamma_2$.

\begin{algorithmic}[1]
    \renewcommand{\algorithmiccomment}[1]{\textcolor{blue}{// #1}}
    
    \STATE \COMMENT{Inversion obtains semantic initial noise. (Step 1)}
    \STATE Perform inversion $\tilde{x}_t = Inv^t_{\theta_{\text{pre-trained}}}(x|\mathbf{c}, \gamma_2)$.

    \STATE \COMMENT{Membership inference attack. (Step 2)}
    \STATE Generate the image $\tilde{x}=G_\theta(\tilde{x}_t|\mathbf{c}, \gamma_1)$.
    \STATE Compute the membership score: $Score=D(x, \tilde{x})$.

    \IF{$Score < \tau$}
        \STATE Conclude that $\mathds{1} = 1$, i.e., $x$ is a member data. 
    \ELSE 
        \STATE Conclude that $\mathds{1} = 0$, i.e., $x$ is a non-member data.
    \ENDIF

\end{algorithmic}
\raggedright \textbf{Output:} Membership status $\mathds{1} \in \{0,1\}$.
\end{algorithm}

\textbf{Step 1}: Given a target image $x_0$ and its corresponding text prompt $\mathbf{c}$, we first employ a pre-trained diffusion model to perform DDIM inversion, thereby obtaining a semantic initial noise $\tilde{x}_t$. 

\textbf{Step 2}: The adversary feeds the semantic initial noise $\tilde{x}_t$ and the same prompt $\mathbf{c}$ into the target model to generate a candidate image $\tilde{x}_0$. If the target image $x_0$ was part of the fine-tuning dataset, the generated candidate image tends to preserve its structural and semantic consistency, resulting in a smaller perceptual distance. Conversely, the non-member samples will yield larger deviations in the generated outputs. The membership inference decision is made by comparing the reconstruction distance using a metric $D(\cdot, \cdot)$. In summary, our method can be summarized as follows:
\begin{equation}
\small
\begin{cases}
\tilde{x}_{t} = Inv^t_{\theta_{pre-trained}}(x_0|\mathbf{c}, \gamma_2), \;\textcolor{red}{(Step\;1)}\\[4pt]
\mathcal{A}(x_i, \theta) = \mathds{1} \left[ D(x_0, G_\theta(\tilde{x}_{t}|\mathbf{c},\gamma_1)) \leq \tau \right]. \; \textcolor{red}{(Step\;2)}
\end{cases}
\end{equation}
where \(D(\cdot, \cdot)\) represents the distance metric. 

\section{Experiments}
\subsection{Experiment Setup}
\label{Setup}
\textbf{Datasets and Models.} We follow the previous stringent assumption that both member and non-member data are drawn from the same distribution \cite{duan2023diffusion, kong2023efficient, zhai2024membership, lian2025unveiling}. We construct member/non-member datasets using 416/417 samples from Pokémon \cite{lambda2023pokemon}, 500/500 samples from text-to-image-2M (T-to-I) \cite{T-to-I}, 2500/2500 samples from MS-COCO \cite{lin2014microsoft}, and 1000/1000 samples from Flickr \cite{young2014image}. We fine-tune Stable Diffusion-v1-4 (SD-v1-4) \cite{compvis2024stablediffusion} using the official fine-tuning scripts from the Hugging-Face Diffusers library \cite{huggingface2024}. 

\vspace{0.3em} 
\noindent\textbf{Evaluation Metrics.} We adopt the evaluation metrics commonly used in prior membership inference attacks for large models \cite{duan2023diffusion, kong2023efficient, pang2023black, he2025towards, fu2024membership}. Specifically, we report the Area Under the Curve (denoted as \textbf{AUC}), which reflects the average success of membership inference attacks. In addition, we measure the True Positive Rate (TPR) at 1\% False Positive Rate (FPR) (\textbf{denoted as T@F=1\%}), which assesses attack efficacy under a strict decision threshold, emphasizing performance at an extremely low FPR. 

\vspace{0.3em} 
\noindent\textbf{Baselines.} We compare our method with existing end-to-end attacks, including NA-P \cite{wu2022membership}, Feature-T \cite{pang2023black}, Feature-D \cite{pang2023black}, Feature-C \cite{pang2023black}, and GD \cite{zhang2024generated}. We also evaluate intermediate result attacks, including SecMI \cite{duan2023diffusion} and PIA \cite{kong2023efficient}.

\vspace{0.3em} 
\noindent\textbf{Implementation Details.} All fine-tuning and inference experiments were conducted on a single RTX 4090 GPU (24 GB). During the DDIM inversion step, we set the guidance scale $\gamma_2 = 1.0$ and the number of steps $i_{\text{step}} = 100$. For the membership inference step, we set the guidance scale $\gamma_1 = 3.5$ and the number of inference steps to $50$. We use the $\ell_{2}$ distance as the $D(\cdot, \cdot)$ by default.

\begin{table*}[ht]
    \centering
    \caption{Hyperparameter analysis of $i_{\text{step}}$ and $\gamma_2$. The results show that our method achieves consistently high performance across a wide range of hyperparameter values, demonstrating its robustness.}
    \label{Hyperparameter}
    \renewcommand{\arraystretch}{1.0} 
    \begin{tabular}{@{}lccccccccccc@{}}
        \toprule       
        \multirow{2}{*}{\makecell{\(i_{step}/\gamma_{2}\)}} & \multicolumn{2}{c}{\(\gamma_{2}=0.0\)} & \multicolumn{2}{c}{\(\gamma_{2}=1.0\)} & \multicolumn{2}{c}{\(\gamma_{2}=3.5\)} & \multicolumn{2}{c}{\(\gamma_{2}=4.5\)} & \multicolumn{2}{c}{\(\gamma_{2}=7.5\)} \\ 
        \cmidrule(lr){2-3}  \cmidrule(lr){4-5} \cmidrule(lr){6-7}  \cmidrule(lr){8-9} \cmidrule(lr){10-11} 
        & AUC & T@F=1\% & AUC & T@F=1\% & AUC & T@F=1\% & AUC & T@F=1\% & AUC & T@F=1\% \\ 
        \midrule
        \(i_{step}=25\) & 89.20 & 21.40 & 89.16 & 21.40 & 87.36 & 21.00 & 87.46 & 21.20 & 87.25 & 21.00 \\ 
        \(i_{step}=50\) & 87.36 & 21.20 & 87.24 & 21.00 & 89.14 & 21.20 & 89.36 & 21.60 & 89.33 & 21.40 \\ 
        \(i_{step}=100\) & 88.48 & 21.00 & 90.46 & 21.80 & 88.56 & 21.00 & 88.61 & 21.00 & 90.12 & 21.80 \\ 
        \(i_{step}=200\) & 89.12 & 21.20 & 89.12 & 21.40 & 91.00 & 22.00 & 91.04 & 22.00 & 89.31 & 21.20 \\ 
        \bottomrule
    \end{tabular}
    \vspace{-5pt}
\end{table*}

\vspace{-5pt}
\subsection{Main Result}
\textbf{Overall Attack Performance.} We report comprehensive attack results and compare the performance of our method against all baselines. As shown in Tab.~\ref{Performance}, our method consistently achieves superior performance across different datasets, delivering significant improvements over existing end-to-end attacks. In particular, compared to the state-of-the-art (SOTA) end-to-end attack, Feature-C, our method can yield improvements of up to 21.77\% in AUC and 11.80\% in TPR@1\%FPR. Notably, Feature-C requires auxiliary data to train shadow models and classifiers, resulting in significant training overhead. In contrast, our method performs the attack solely through threshold setting, achieving superior performance with a lower computational cost. Moreover, our method remains highly competitive compared to intermediate result attacks, e.g., SecMI and PIA, and outperforms them on the MS-COCO dataset, all while not requiring access to the intermediate results.

\begin{table}
    \centering
    \caption{Ablation study on different datasets, showing the impact of semantic initial noise on attack performance.}
    \vspace{-3pt}
    \label{ablation}
    \renewcommand{\arraystretch}{1.0} 
    \resizebox{0.48\textwidth}{!}{
    \begin{tabular}{@{}lccccccccc@{}}
        \toprule
        \multirow{2}{*}{\makecell{Method}} & \multicolumn{2}{c}{Pokémon} & \multicolumn{2}{c}{T-to-I} & \multicolumn{2}{c}{MS-COCO} & \multicolumn{2}{c}{Flickr} \\ 
        \cmidrule(lr){2-3} \cmidrule(lr){4-5} \cmidrule(lr){6-7} \cmidrule(lr){8-9} 
        & AUC & T@F=1\% & AUC & T@F=1\% & AUC & T@F=1\% & AUC & T@F=1\% \\ 
        \midrule
        Naive & 55.66 & 6.50 & 72.76 & 10.50 & 62.24 & 7.00 & 57.29 & 5.00 \\ 
        Ours & 82.44 & 14.00 & 89.24 & 21.60 & 90.46 & 21.80 & 76.23 & 16.00 \\ 
        \rowcolor[gray]{0.9}
        \textbf{Gain} & \textbf{+26.78} & \textbf{+7.50} & \textbf{+16.48} & \textbf{+11.10} & \textbf{+28.22} & \textbf{+14.80} & \textbf{+18.94} & \textbf{+11.00} \\ 
        \bottomrule
    \end{tabular}
    }
    \vspace{-5pt}
\end{table}

\begin{figure}[h]
    \centering
    \vspace{-5pt}
    \begin{subfigure}{0.32\linewidth}
        \includegraphics[width=\linewidth]{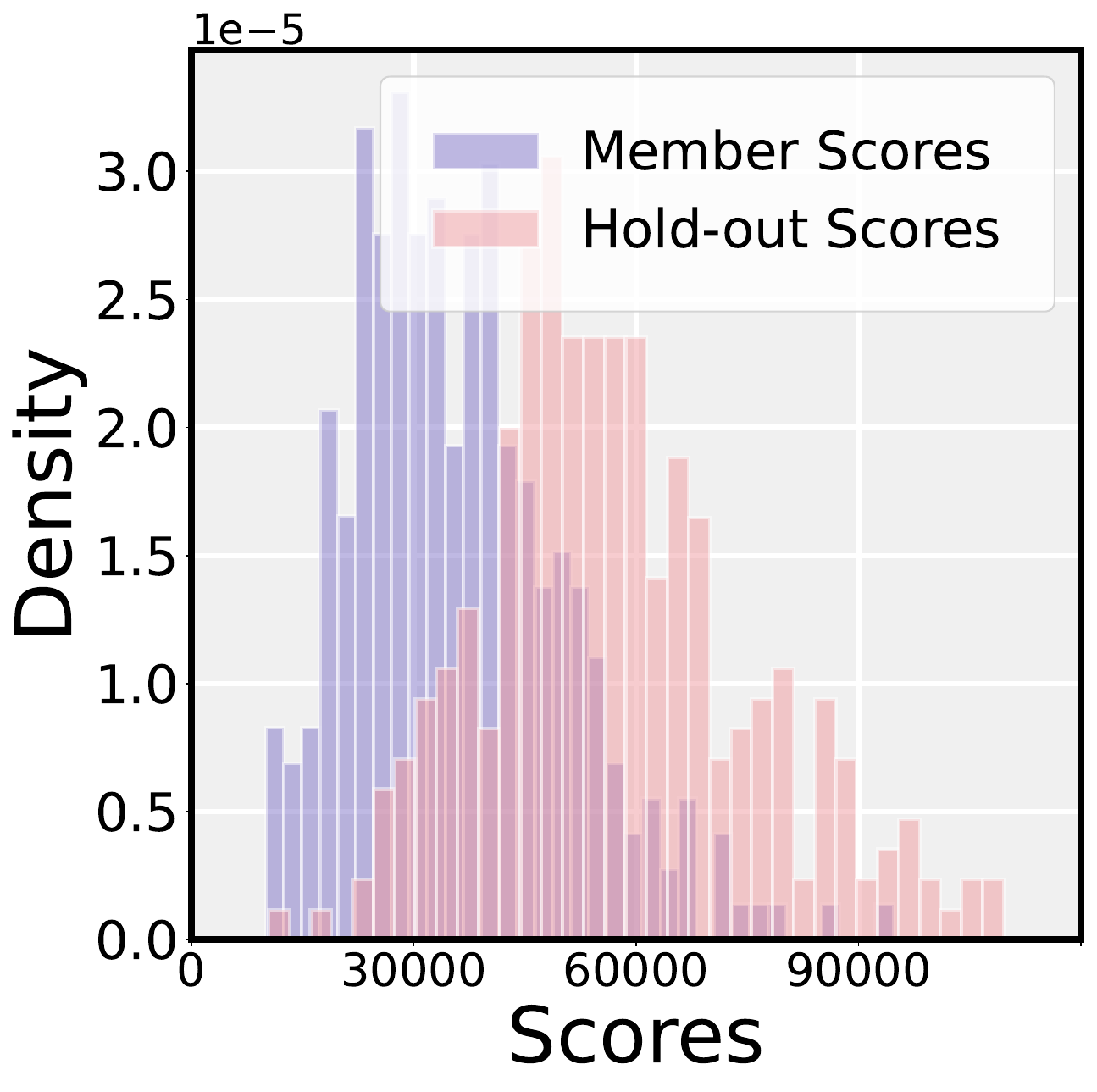}
    \end{subfigure}
    \hfill
    \begin{subfigure}{0.32\linewidth}
        \includegraphics[width=\linewidth]{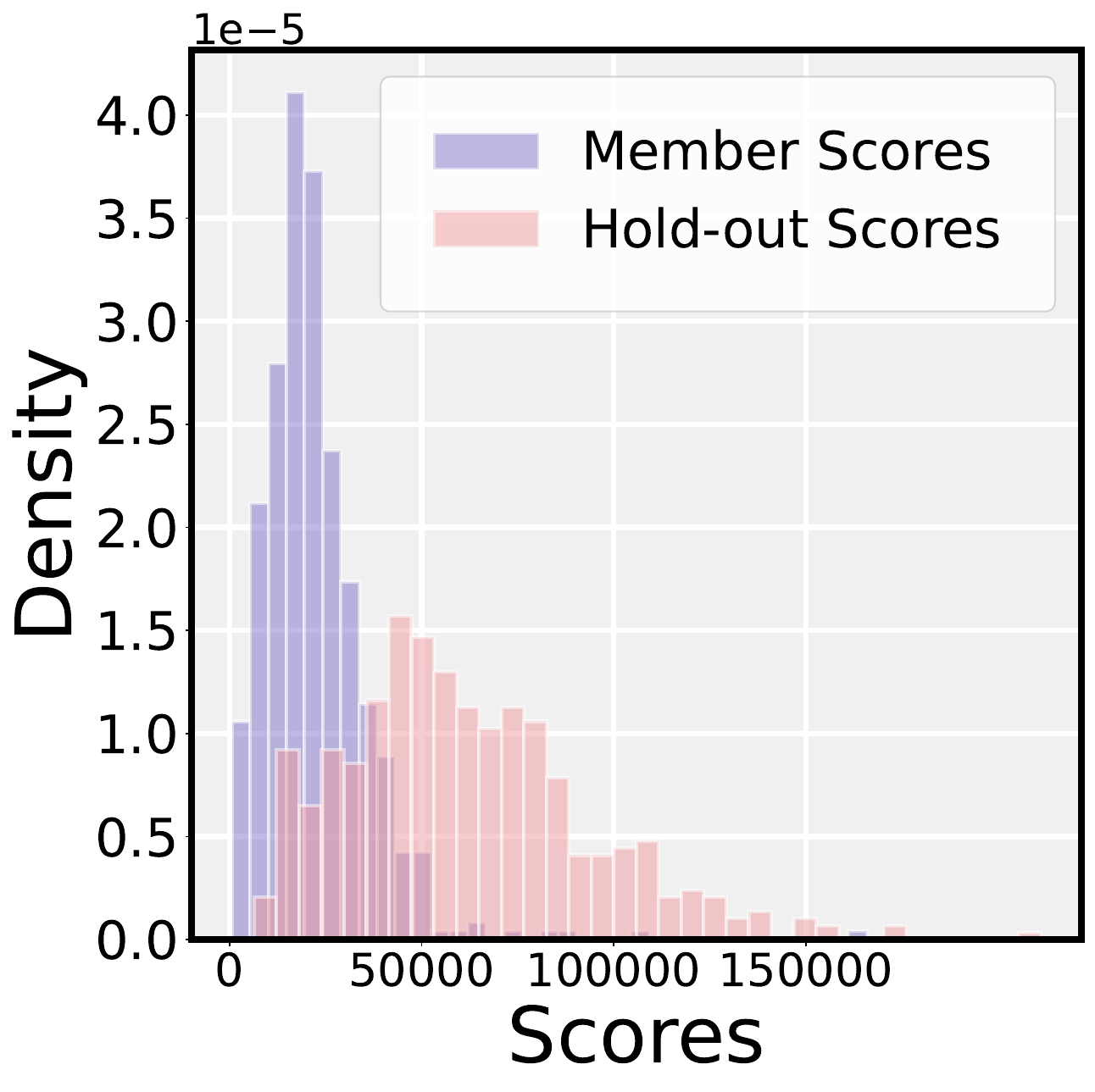}
    \end{subfigure}
    \hfill
    \begin{subfigure}{0.32\linewidth}
        \includegraphics[width=\linewidth]{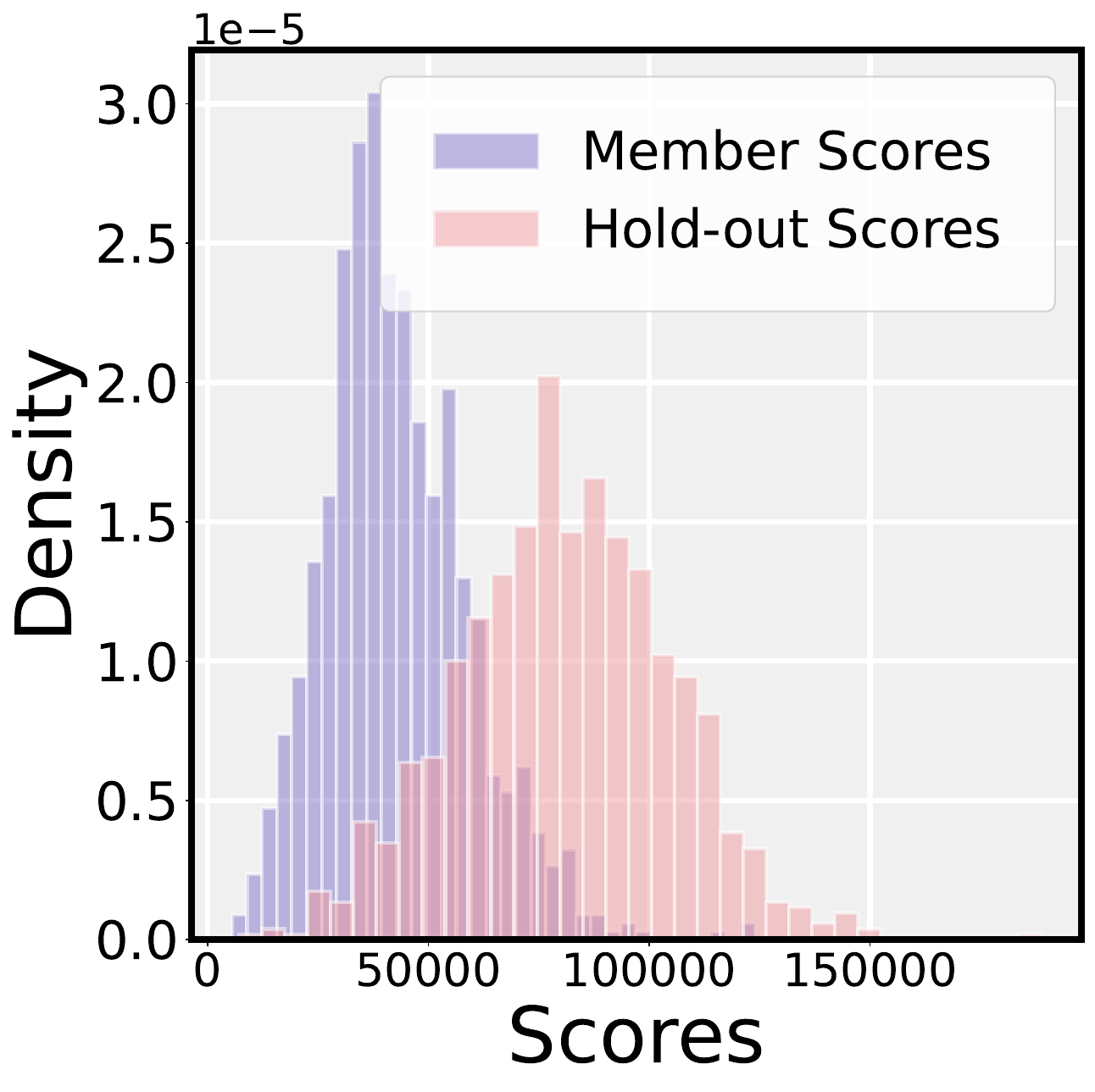}
    \end{subfigure}
    
    Membership score distribution of our method.
    \vspace{0.8em}
    
    \begin{subfigure}{0.32\linewidth}
        \includegraphics[width=\linewidth]{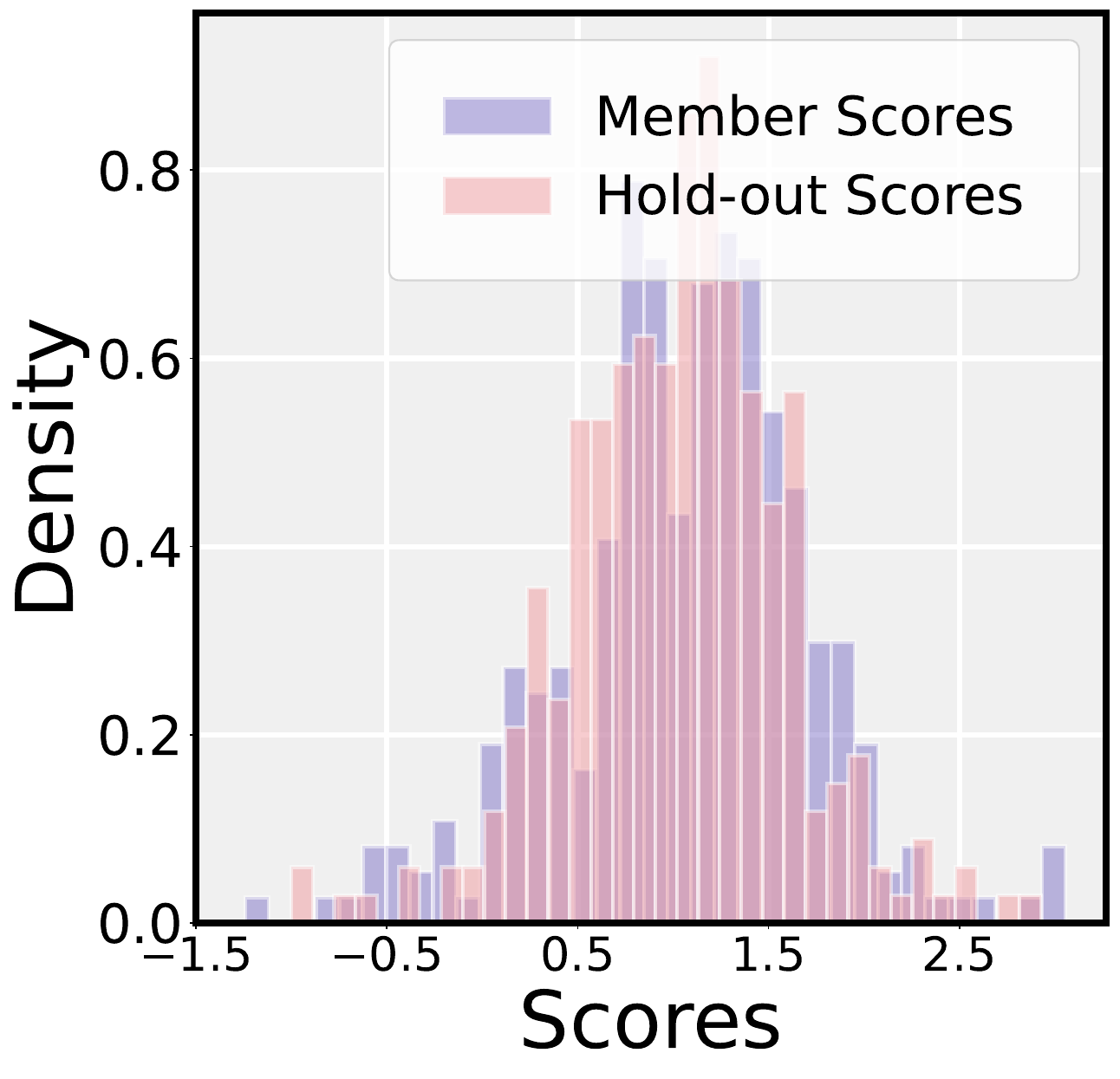}
    \end{subfigure}
    \hfill
    \begin{subfigure}{0.32\linewidth}
        \includegraphics[width=\linewidth]{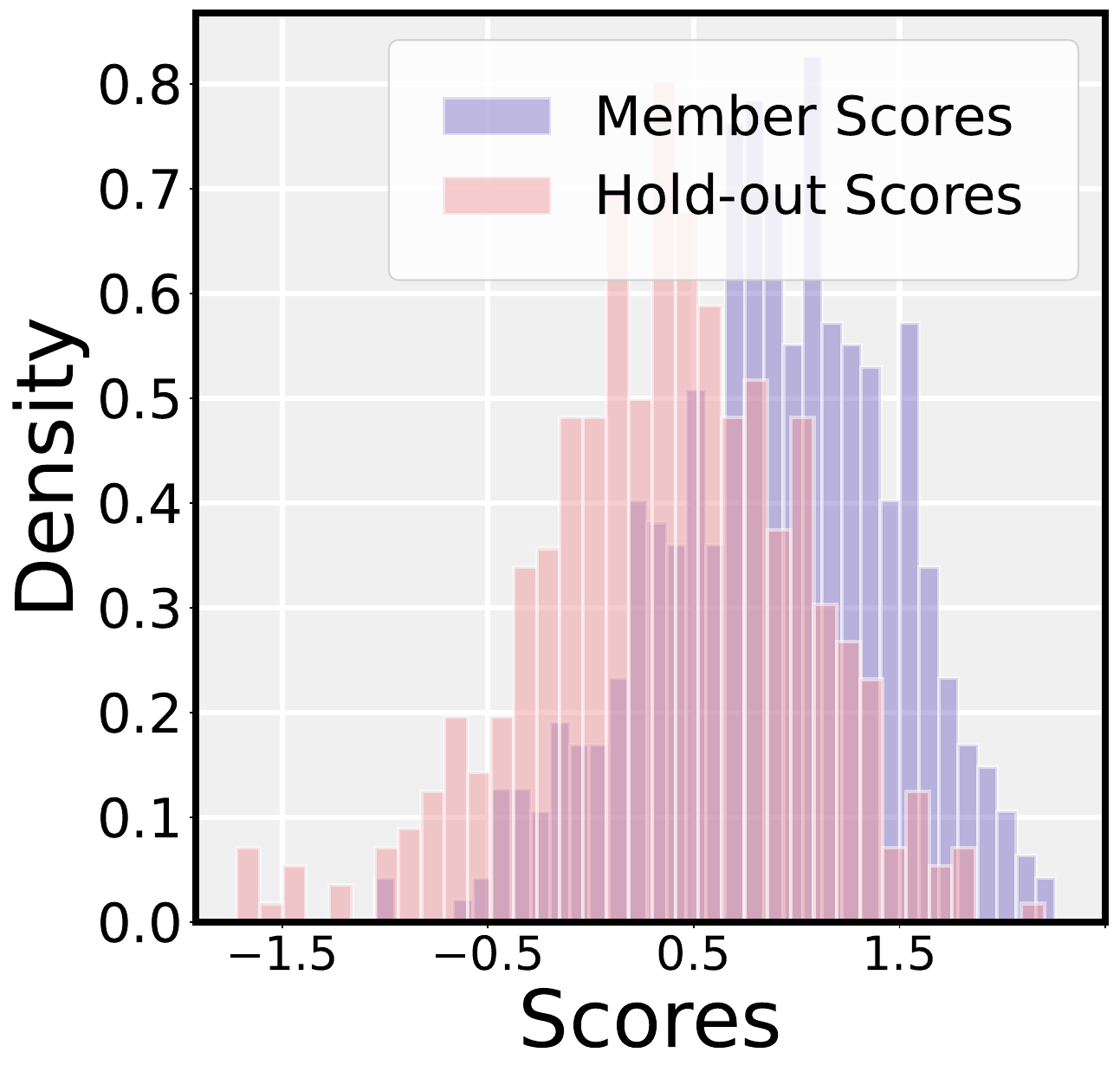}
    \end{subfigure}
    \hfill
    \begin{subfigure}{0.32\linewidth}
        \includegraphics[width=\linewidth]{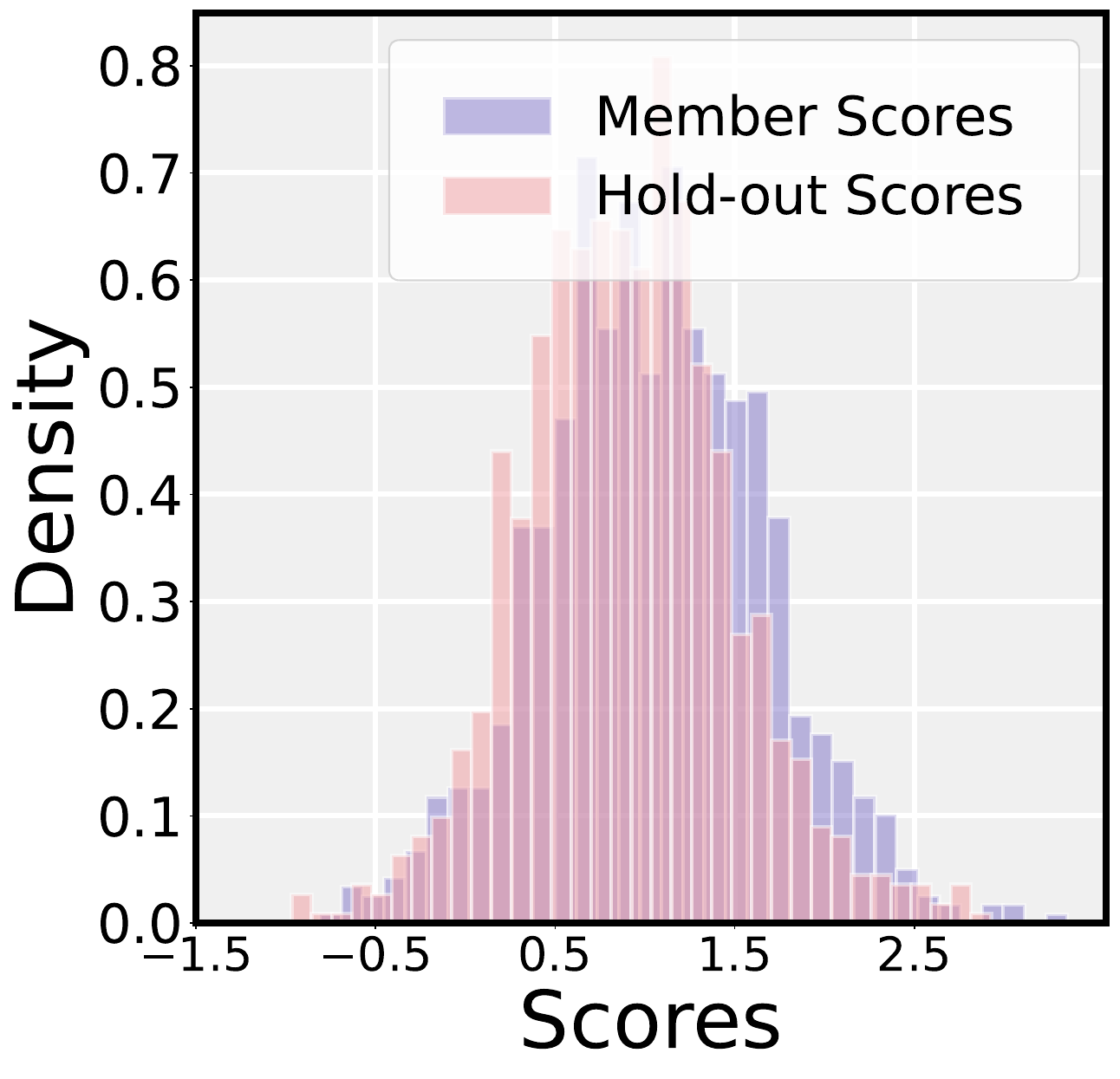}
    \end{subfigure}

    \vspace{0.4em}
    Membership score distribution of Feature-T.
    \vspace{-3pt}
    \caption{Membership score distribution of member and non-member data in the Pokémon, T-to-I, and MS-COCO dataset, arranged from left to right. The score distribution gap between member data and hold-out data is significantly larger in our method.}
    \label{Score_distribution}

\end{figure}

\vspace{0.3em} 
\noindent \textbf{Visual Analysis of Attack Performance.} To further illustrate the effectiveness of our method, we compare it against Feature-T, the previous most potent threshold-based end-to-end attack. As shown in Fig.~\ref{Score_distribution}, we visualize the membership score distributions of member and non-member data. The separation between the two distributions achieved by our method is substantially larger than that of Feature-T. This visualization provides intuitive evidence of the enhanced distinguishability achieved by our method. 

\vspace{-3pt}
\subsection{Ablation Study}
\textbf{Contribution of Semantic Initial Noise}. In this section, we define the Naive method, which generates images solely from the given caption using randomly initialized noise. The distance between the generated image and the target image is then measured to determine membership. The ablation results are reported in Tab.~\ref{ablation}. On average, the AUC improved by 21.57\% across different datasets, and T@F=1\% increased by 10.63\%. It clearly shows that injecting semantics into the initial noise substantially improves the attack performance. These results validate the effectiveness of our approach and underscore the importance of carefully managing and protecting initial noise.
\vspace{0.2em} \\
\textbf{Ablation on Hyperparameter.} We validate the impact of hyperparameters \(i_{step}\) and \(\gamma_{2}\) on the performance of our method. As shown in Tab.~\ref{Hyperparameter}, our method demonstrates very low sensitivity to \(i_{step}\) and \(\gamma_{2}\), with the best and worst AUC being 91.04\% and 87.25\%, respectively (a variation of only 3.79\%). It provides strong confirmation of the robustness of our method.

\subsection{A More Knowledge-Restricted Adversary}
\label{Restricted}
\textbf{Without Access to the Model Architecture.} In real-world scenarios, model publishers may deliberately withhold the architecture version of the pre-trained model used during fine-tuning, thereby increasing the difficulty of attacks. To evaluate performance under this condition, we use SD-v1-5 \cite{runwayml2024}, SD-2-1 \cite{sd2.1}, SDXL-turbo \cite{sdxl}, and Dreamshaper-XL \cite{dreamshaper-xl} to perform inversion and obtain initial noise, where the architectural discrepancy from the SD-v1-4 gradually increases. And these models are used as shadow models for the baseline attacks. As shown in Tab.~\ref{Different}, our method remains effective even when the architecture version of the target model is unknown. Although all methods exhibit performance degradation as the architectural gap widens, our method achieves the best performance even using SDXL-turbo and Dreamshaper-XL, whose architectures differ substantially from the target model. This finding aligns with the observations in \cite{wang2024silent}, which indicate that the semantics of the initial noise can be transferred across models due to the shared distributions learned during large-scale pre-training. This further validates that semantic initial noise can be leveraged to reveal membership information.

\begin{table}[h]
    \centering
    \caption{Attack performance using semantic initial noise obtained from different models, demonstrating that our method remains effective when the target model's architecture version is unknown. Intermediate result attacks are not applicable in this scenario.}
    \vspace{-3pt}
    \label{Different}
    \resizebox{0.48\textwidth}{!}{
    \begin{tabular}{@{}lccccccccc@{}}
        \toprule
        \multirow{2}{*}{\makecell{Method}} & \multicolumn{2}{c}{SD-v1-5} & \multicolumn{2}{c}{SD-2-1} & \multicolumn{2}{c}{SDXL-turbo} & \multicolumn{2}{c}{Dreamshaper}  \\ 
        \cmidrule(lr){2-3} \cmidrule(lr){4-5} \cmidrule(lr){6-7} \cmidrule(lr){8-9} 
        & AUC & T@F=1\% & AUC & T@F=1\% & AUC & T@F=1\% & AUC & T@F=1\% \\ 
        \midrule
        NA-P & 70.53 & 6.60 & 69.21 & 6.60 & 55.53 & 2.20 & 56.02 & 2.40 \\ 
        Feature-C & 83.41 & 17.40 & 81.28 & 16.70 & 68.41 & 6.20 & 67.42 & 6.00 \\
        Feature-D & 65.00 & 6.00 & 64.59 & 6.00 & 58.00 & 4.00 & 57.66 & 4.20 \\
        \rowcolor[gray]{0.9}
        \textbf{Ours} & \textbf{89.03} & \textbf{20.20} & \textbf{88.04} & \textbf{19.60} & \textbf{76.91} & \textbf{8.40} & \textbf{76.87} & \textbf{8.20} \\
        \bottomrule
    \end{tabular}
    }
    \vspace{-5pt}
\end{table}

\begin{table*}[t]
    \centering
    \caption{Attack performance under defense mechanisms. All attacks experience
    varying degrees of performance degradation, while our method consistently
    achieves the best performance.}
    \label{defense}
    \vspace{-5pt}

    \begingroup
    \setlength{\tabcolsep}{1.2pt}
    \renewcommand{\arraystretch}{1.08}

    \begin{tabular}{@{}cc*{8}{cc}@{}}
        \toprule
        \multirow{2}{*}{\(SS_{e_i}\)}
        & \multirow{2}{*}{DataAug}
        & \multicolumn{2}{c}{SecMI}
        & \multicolumn{2}{c}{PIA}
        & \multicolumn{2}{c}{NA-P}
        & \multicolumn{2}{c}{GD}
        & \multicolumn{2}{c}{Feature-T}
        & \multicolumn{2}{c}{Feature-C}
        & \multicolumn{2}{c}{Feature-D}
        & \multicolumn{2}{c}{\textbf{Ours}} \\
        \cmidrule(lr){3-4}
        \cmidrule(lr){5-6}
        \cmidrule(lr){7-8}
        \cmidrule(lr){9-10}
        \cmidrule(lr){11-12}
        \cmidrule(lr){13-14}
        \cmidrule(lr){15-16}
        \cmidrule(l){17-18}

        & & AUC & T@F=1\%
          & AUC & T@F=1\%
          & AUC & T@F=1\%
          & AUC & T@F=1\%
          & AUC & T@F=1\%
          & AUC & T@F=1\%
          & AUC & T@F=1\%
          & \textbf{AUC} & \textbf{T@F=1\%} \\
        \midrule

        $\times$ & $\times$
        & 89.43 & 16.99
        & 74.36 & 6.40
        & 64.77 & 5.57
        & 51.31 & 1.00
        & 59.43 & 4.00
        & 73.26 & 14.60
        & 59.01 & 3.60
        & \textbf{91.12} & \textbf{22.90} \\

        $\times$ & $\checkmark$
        & 89.37 & 16.79
        & 71.38 & 5.20
        & 63.41 & 5.20
        & 51.22 & 1.00
        & 57.70 & 3.20
        & 73.08 & 14.60
        & 58.00 & 3.20
        & \textbf{90.46} & \textbf{21.80} \\

        $\checkmark$ & $\times$
        & 52.11 & 2.10
        & 58.93 & 3.20
        & 62.98 & 4.00
        & 51.05 & 1.00
        & 57.53 & 3.20
        & 71.99 & 13.40
        & 57.20 & 3.00
        & \textbf{87.68} & \textbf{18.00} \\

        $\checkmark$ & $\checkmark$
        & 51.21 & 1.90
        & 54.41 & 1.50
        & 62.56 & 4.00
        & 51.03 & 1.00
        & 57.20 & 3.20
        & 71.02 & 13.20
        & 57.50 & 2.60
        & \textbf{86.74} & \textbf{17.90} \\

        \bottomrule
    \end{tabular}
    \endgroup

    \vspace{-7pt}
\end{table*}

\noindent \textbf{Lacking Access to Image Captions.} In reality, attackers may not have the image captions used for fine-tuning. Therefore, we also evaluate the attack when the image captions are unavailable. To address this, we employ BLIP \cite{li2022blip} to generate captions for the images and use these generated captions to conduct the attack. As shown in Tab.~\ref{blip}, using the MS-COCO dataset as an example, our method outperforms the strongest competitor, Feature-C, with an AUC improvement of 10.66\% and a T@F=1\% improvement of 2.20\%. The experiments reveal that end-to-end attacks are vulnerable to the absence of captions, leading to a noticeable degradation in performance. This is likely because end-to-end attacks involve a multi-step prediction process during image generation, making them more sensitive to changes in the caption. Although our method is also affected under this setting and exhibits a performance drop, it still achieves the best results in end-to-end attacks.

\begin{table}[h]
    \centering
    \caption{Attack performance using BLIP-generated captions. Our method achieves the best performance.}
    \vspace{-3pt}
    \label{blip}
    \renewcommand{\arraystretch}{1.0} 
    \resizebox{0.48\textwidth}{!}{
    \begin{tabular}{@{}lccccccc@{}}
        \toprule
        \multirow{2}{*}{\makecell{Method}} & \multicolumn{2}{c}{T-to-I} & \multicolumn{2}{c}{MS-COCO} & \multicolumn{2}{c}{Flickr} \\ 
        \cmidrule(lr){2-3} \cmidrule(lr){4-5} \cmidrule(lr){6-7}  
        & AUC & T@F=1\% & AUC & T@F=1\% & AUC & T@F=1\% \\ 
        \midrule
        NA-P & 54.03 & 1.00 & 53.00 & 1.00 & 52.01 & 1.00 \\
        GD & 52.23 & 1.00 & 51.03 & 1.00 & 50.71 & 1.00 \\
        Feature-T & 55.80 & 1.00 & 50.20 & 1.00 & 50.10 & 1.00 \\
        Feature-C & 63.20 & 4.00 & 52.88 & 3.40 & 50.98 & 1.00 \\
        Feature-D & 53.00 & 2.00 & 50.20 & 1.00 & 50.20 & 1.00 \\
        \rowcolor[gray]{0.9}
        \textbf{Ours} & \textbf{71.47} & \textbf{6.80} & \textbf{63.54} & \textbf{5.60} & \textbf{57.82} & \textbf{4.80} & \\   
        \bottomrule
    \end{tabular}
    }
    \vspace{-7pt}
\end{table}

\subsection{Impact of Defense}
To assess the robustness of our method, we investigate the impact of the SOTA defense method \(SS_{e_i}\) \cite{wen2024detecting}. This defense dynamically evaluates the model's memorization during training and adjusts the training process accordingly. Following this method, we set the \(SS_{e_i}\) threshold to \(4\). Additionally, data augmentation techniques are commonly employed to mitigate MIAs. During the fine-tuning process of the diffusion model, Random-Crop and Random-Flip are applied by default \citep{huggingface2024}. Following previous work \cite{duan2023diffusion, pang2023white, zhai2024membership}, we also investigate the impact of data augmentation on the performance of attacks. We conduct an in-depth analysis of the performance changes of various attacks before and after applying defenses on the MS-COCO dataset, as shown in Tab.~\ref{defense}. Experimental results show that our method maintains excellent performance even against the advanced defense strategies. In the presence of both defense mechanisms, our method achieves substantially better performance than all the other methods. Specifically, the intermediate result attacks show a significant performance drop under defense. In our method, AUC and TPR@1\%FPR decrease by only 4.38\% and 5.00\%, respectively, compared to performance without any defense. 

\vspace{-2pt}
\subsection{Visualization of Generated Results.}
\begin{figure}[h]
    \centering
    \setlength{\tabcolsep}{1pt} 
    \renewcommand{\arraystretch}{0.5} 
    \vspace{-5pt}
     \begin{tabular}{cccccc}
        \multicolumn{3}{c}{Member} & \multicolumn{3}{c}{Non-member} \\[2pt]
        \cmidrule(lr){0-2} \cmidrule(lr){4-6}
        Original & Naive & Ours & Original & Naive & Ours \\
        \includegraphics[width=0.16\linewidth]{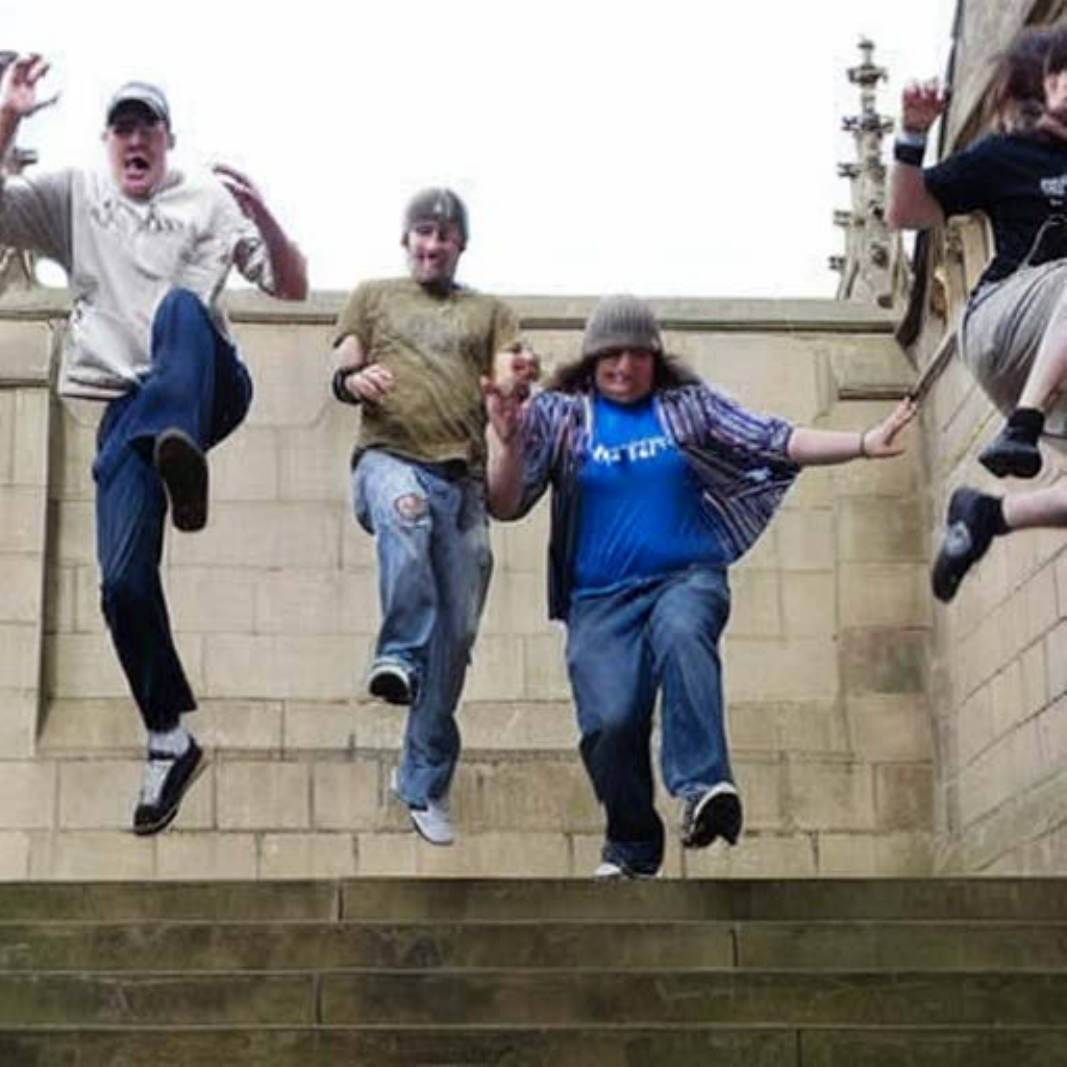} &
        \includegraphics[width=0.16\linewidth]{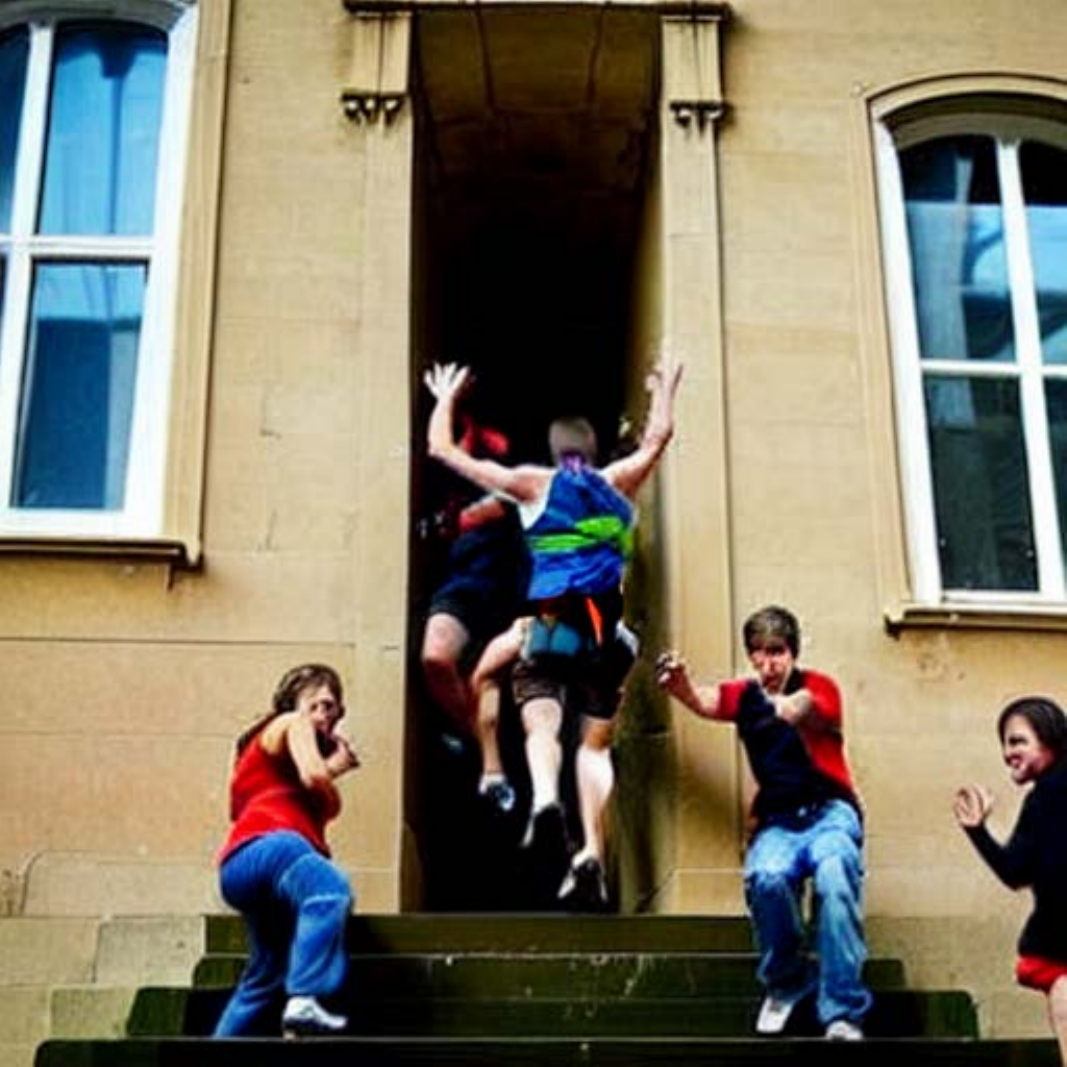} &
        \includegraphics[width=0.16\linewidth]{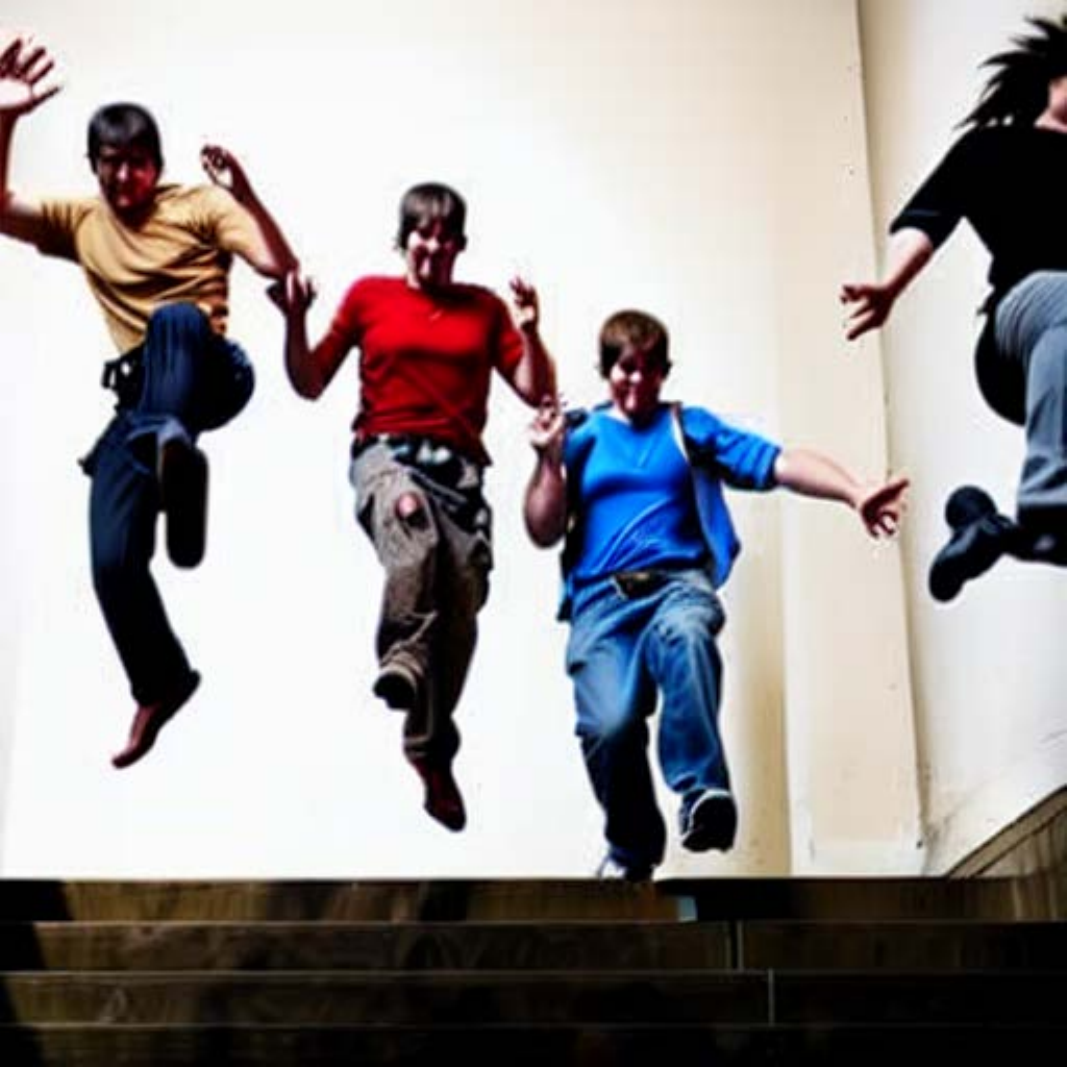} &
        \includegraphics[width=0.16\linewidth]{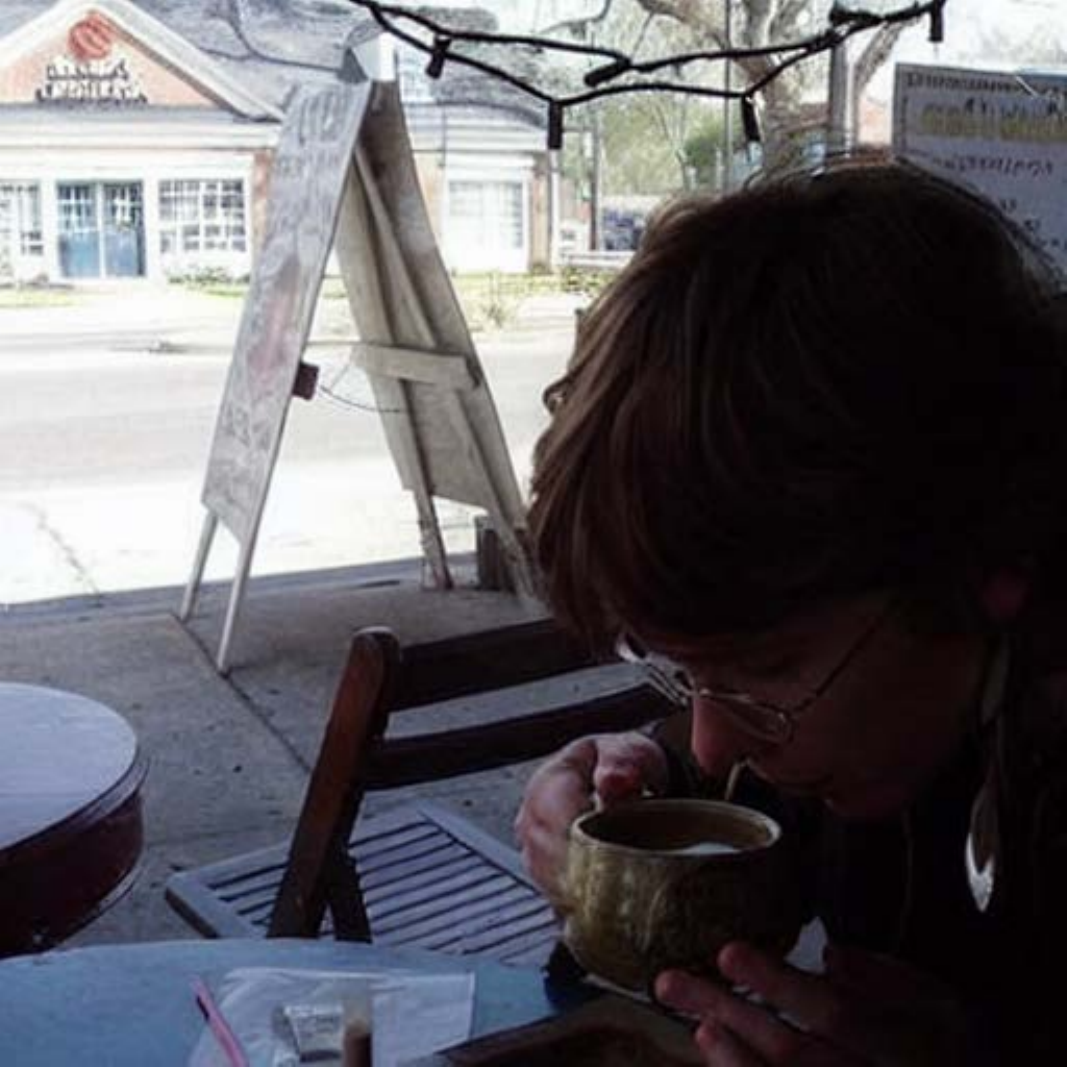} &
        \includegraphics[width=0.16\linewidth]{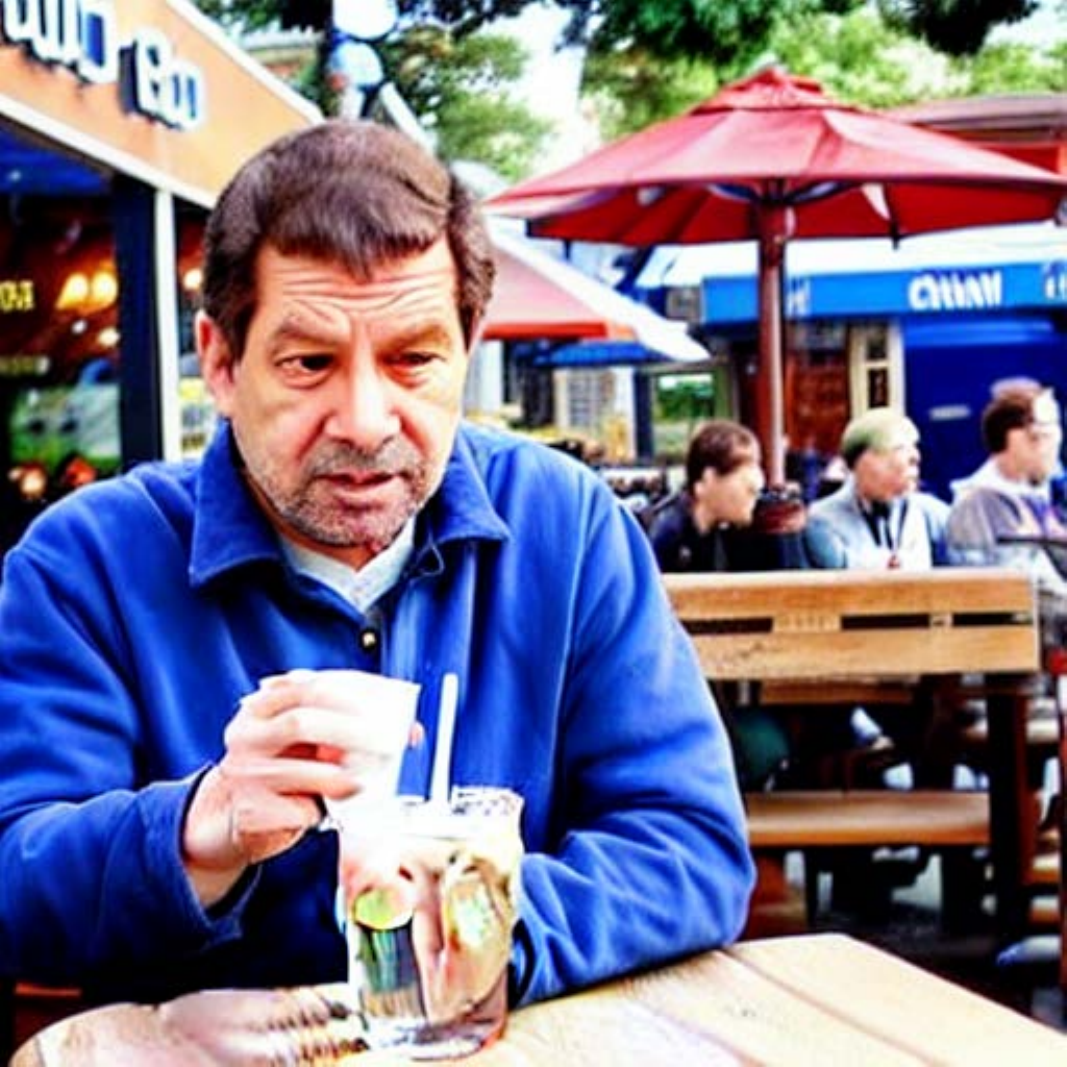} &
        \includegraphics[width=0.16\linewidth]{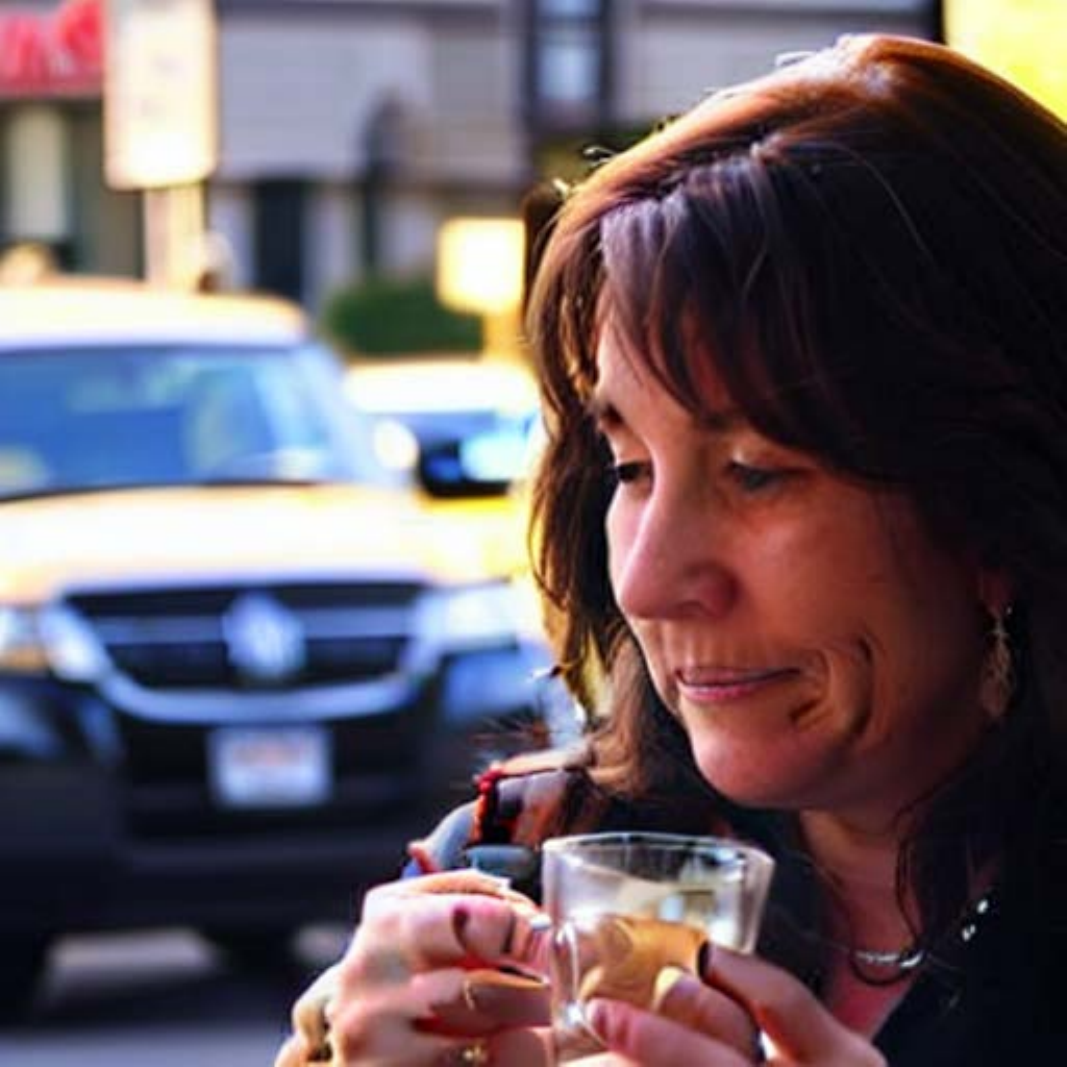} \\

        \includegraphics[width=0.16\linewidth]{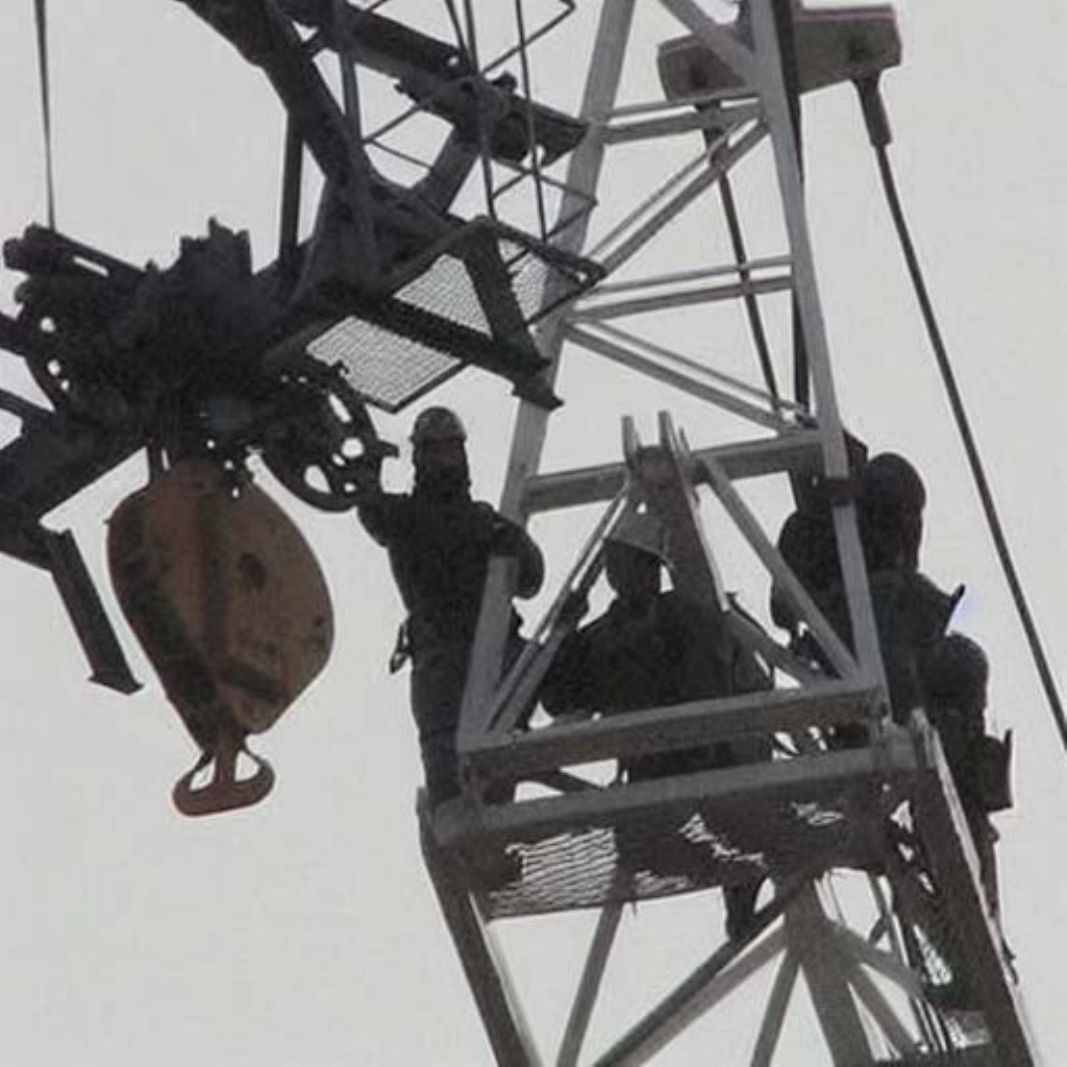} &
        \includegraphics[width=0.16\linewidth]{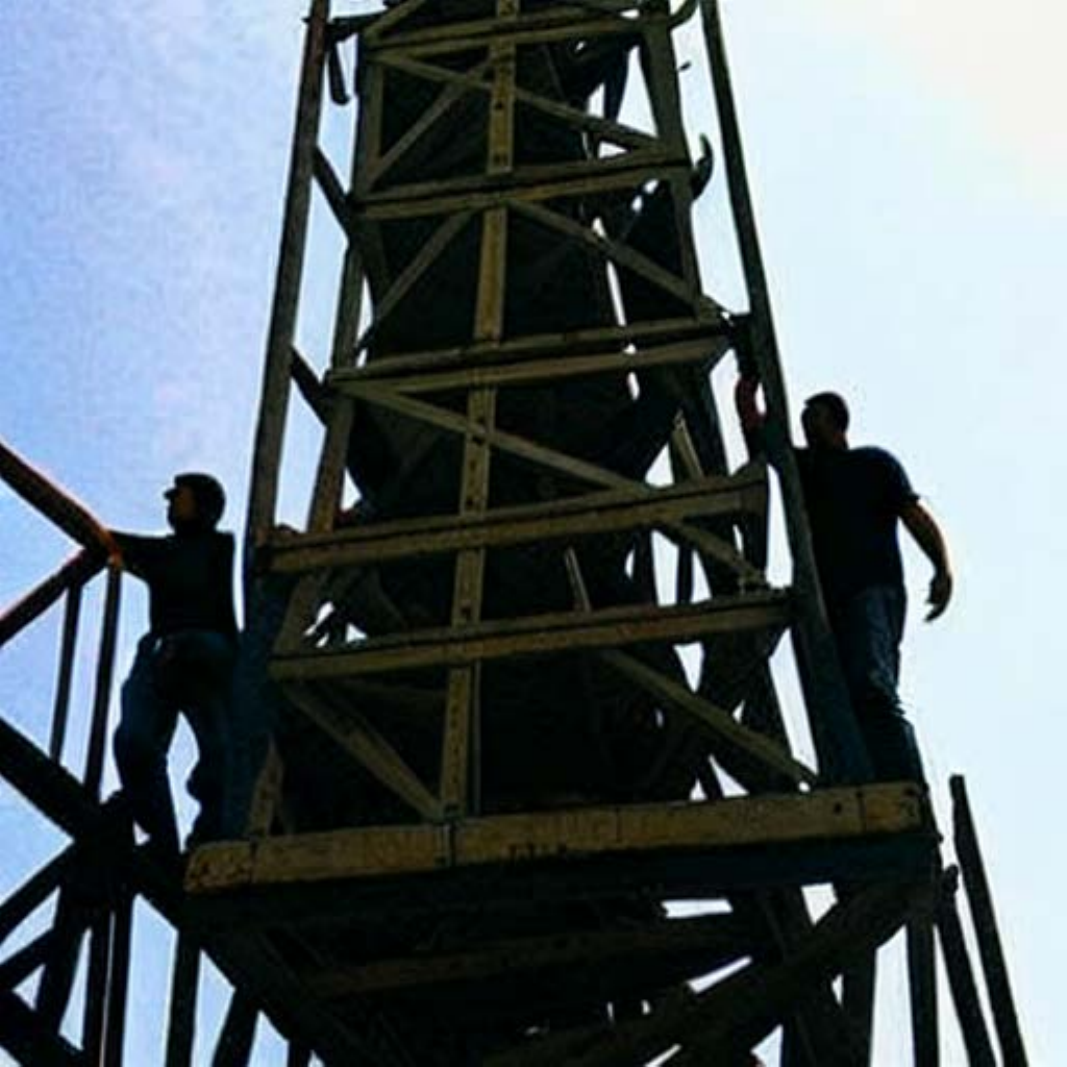} &
        \includegraphics[width=0.16\linewidth]{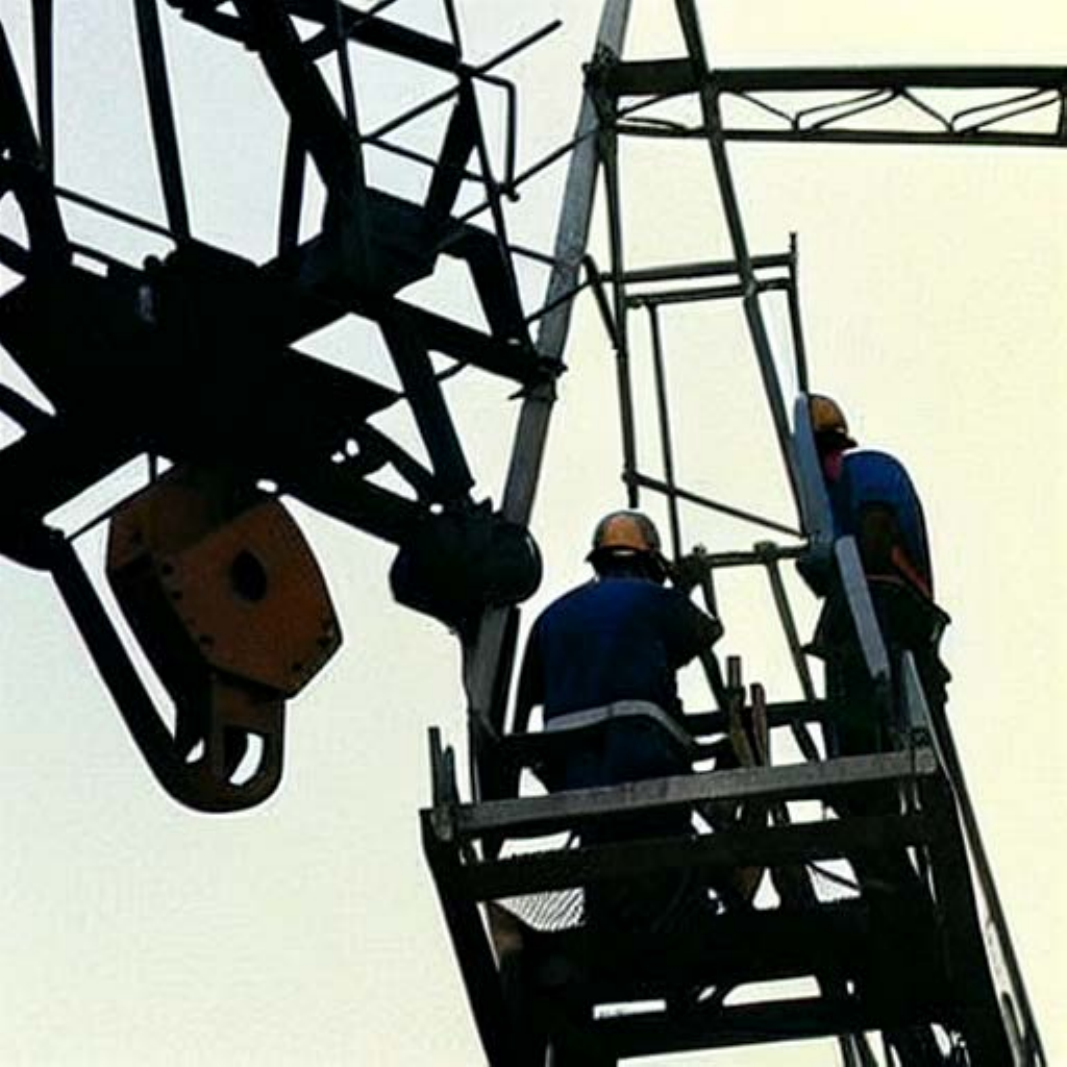} &
        \includegraphics[width=0.16\linewidth]{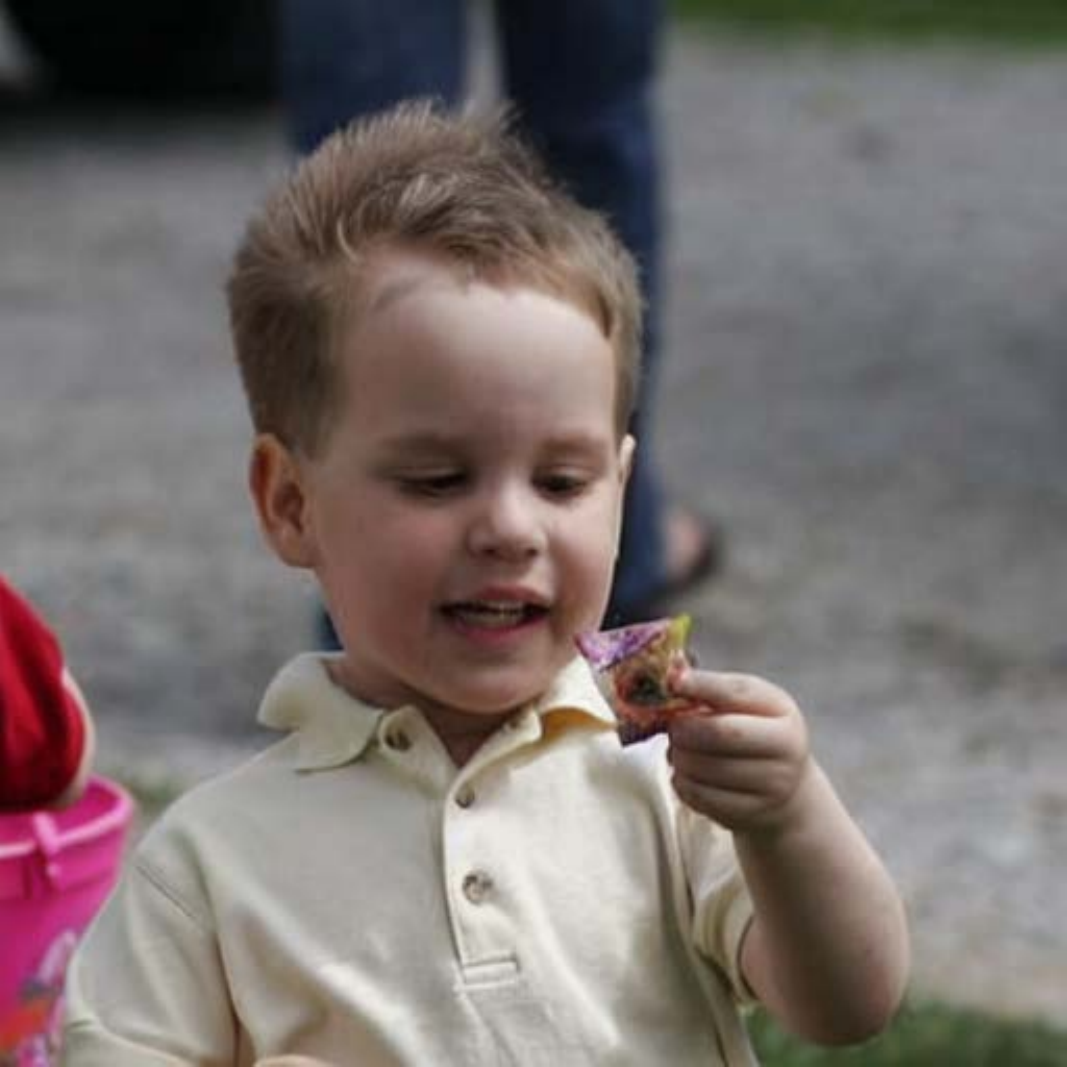} &
        \includegraphics[width=0.16\linewidth]{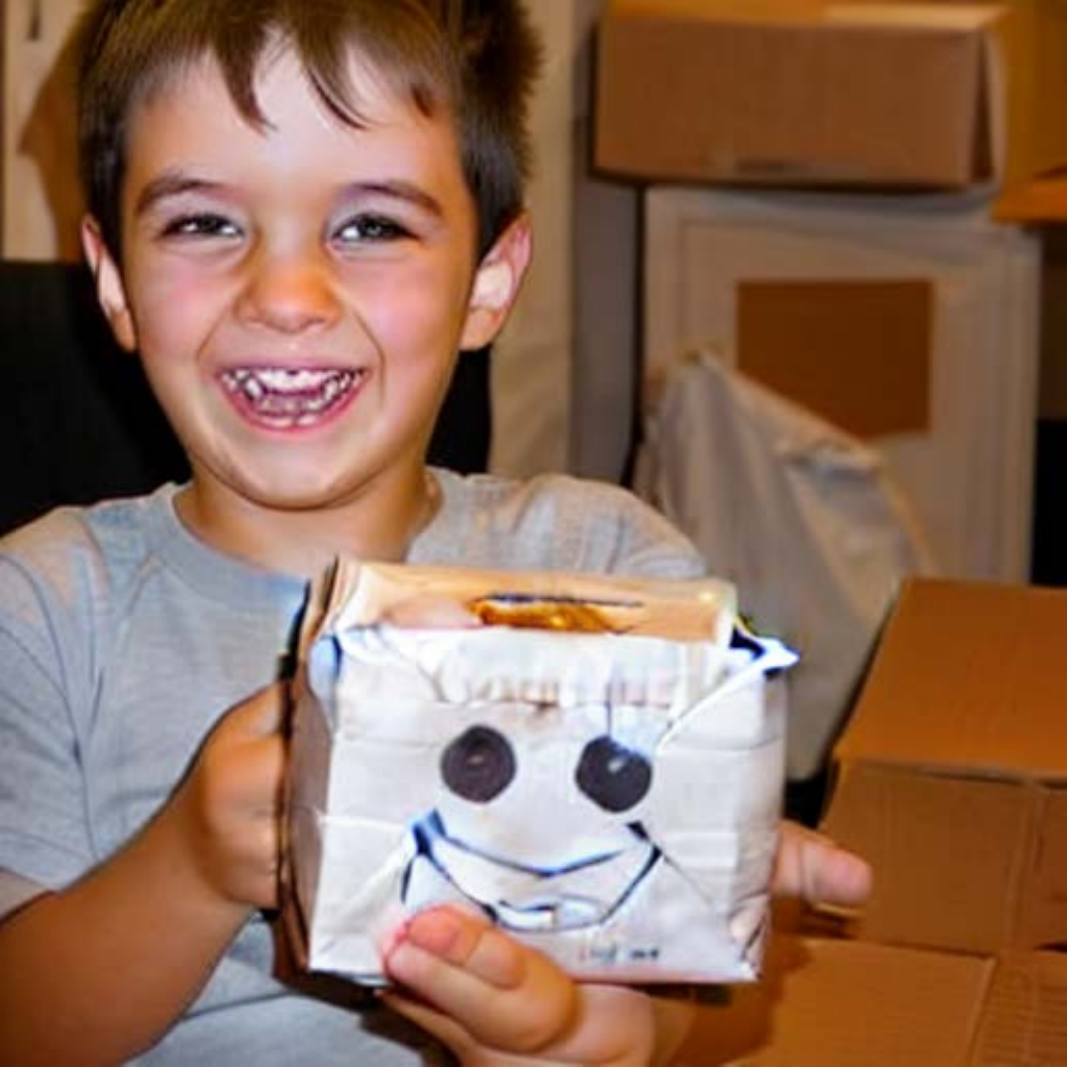} &
        \includegraphics[width=0.16\linewidth]{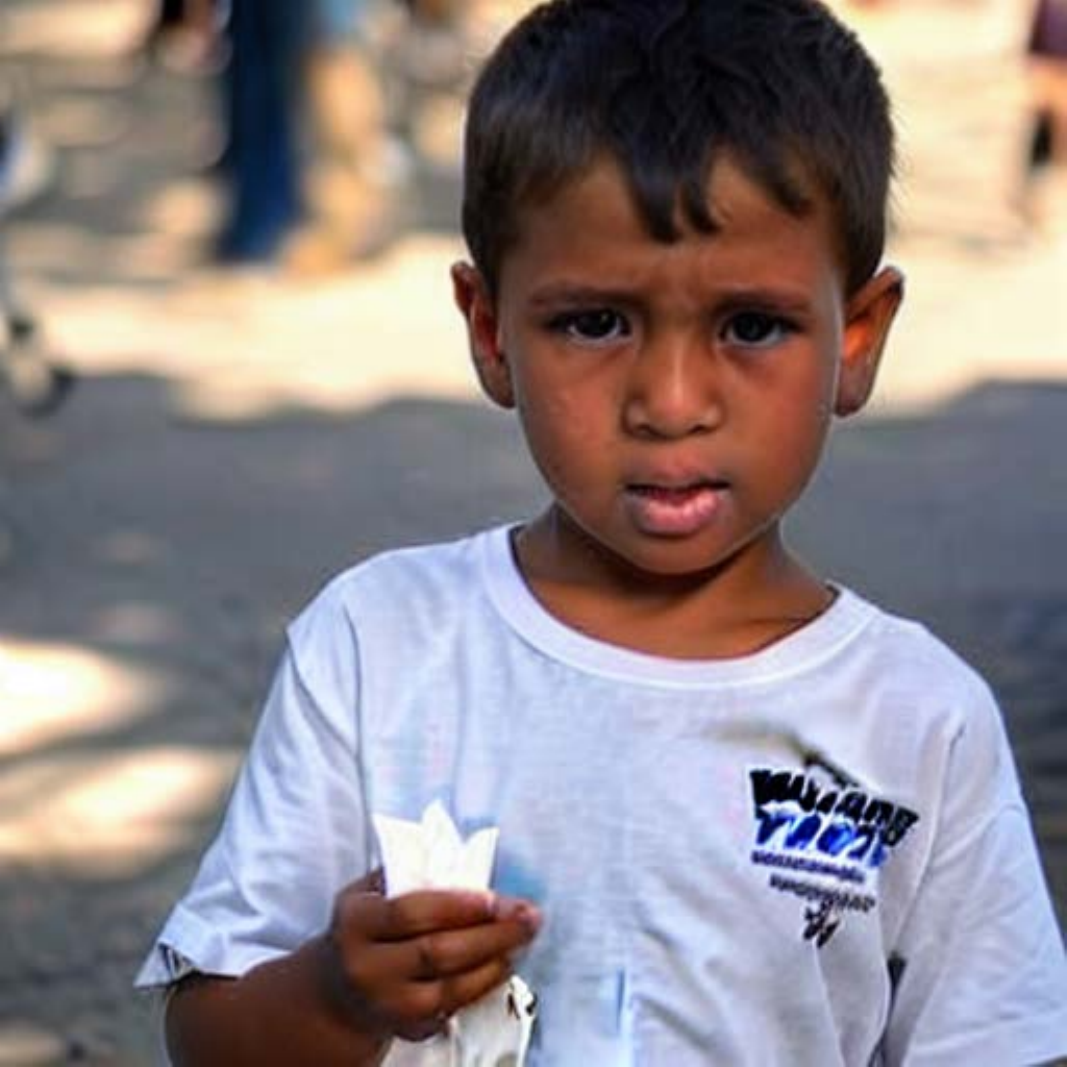} \\

        \includegraphics[width=0.16\linewidth]{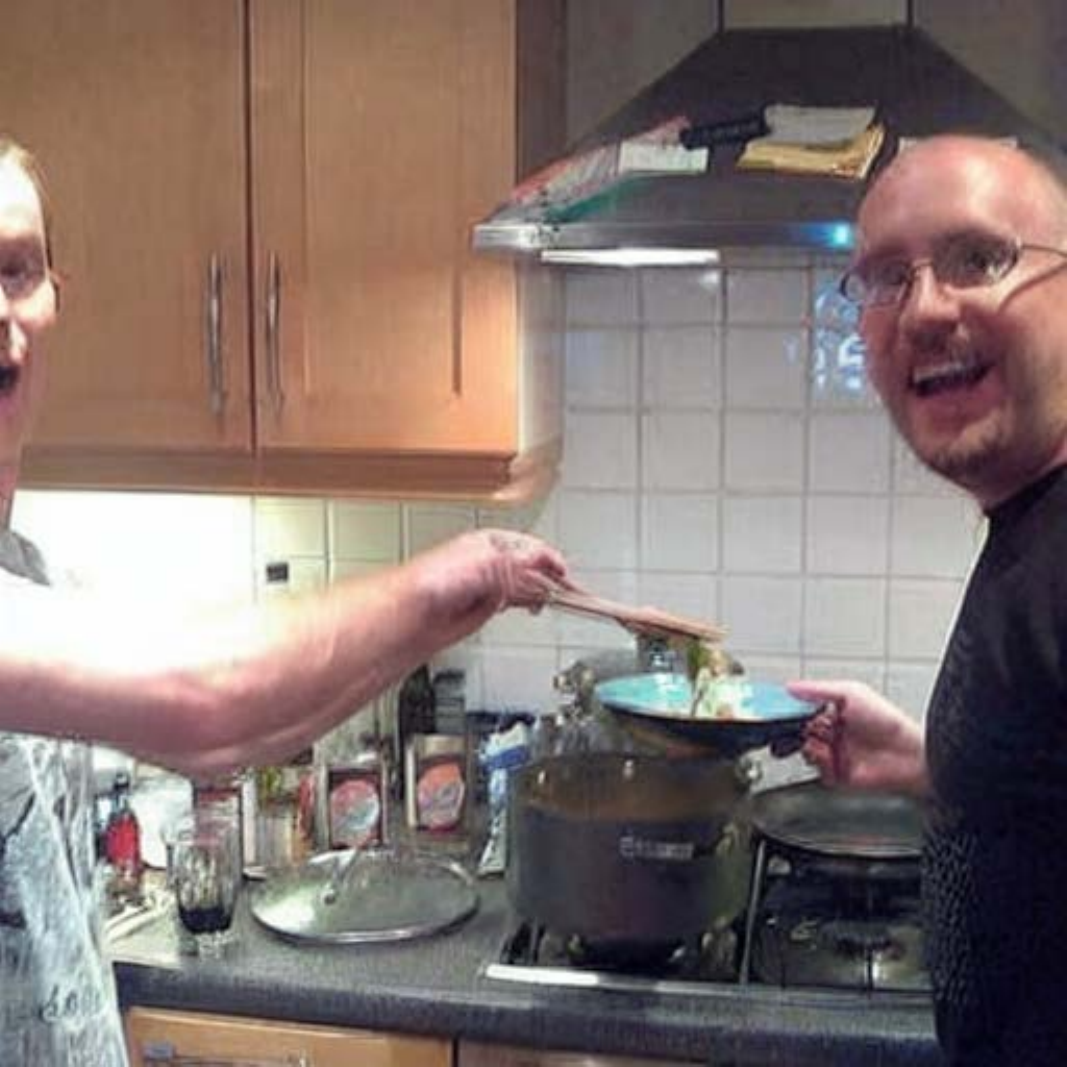} &
        \includegraphics[width=0.16\linewidth]{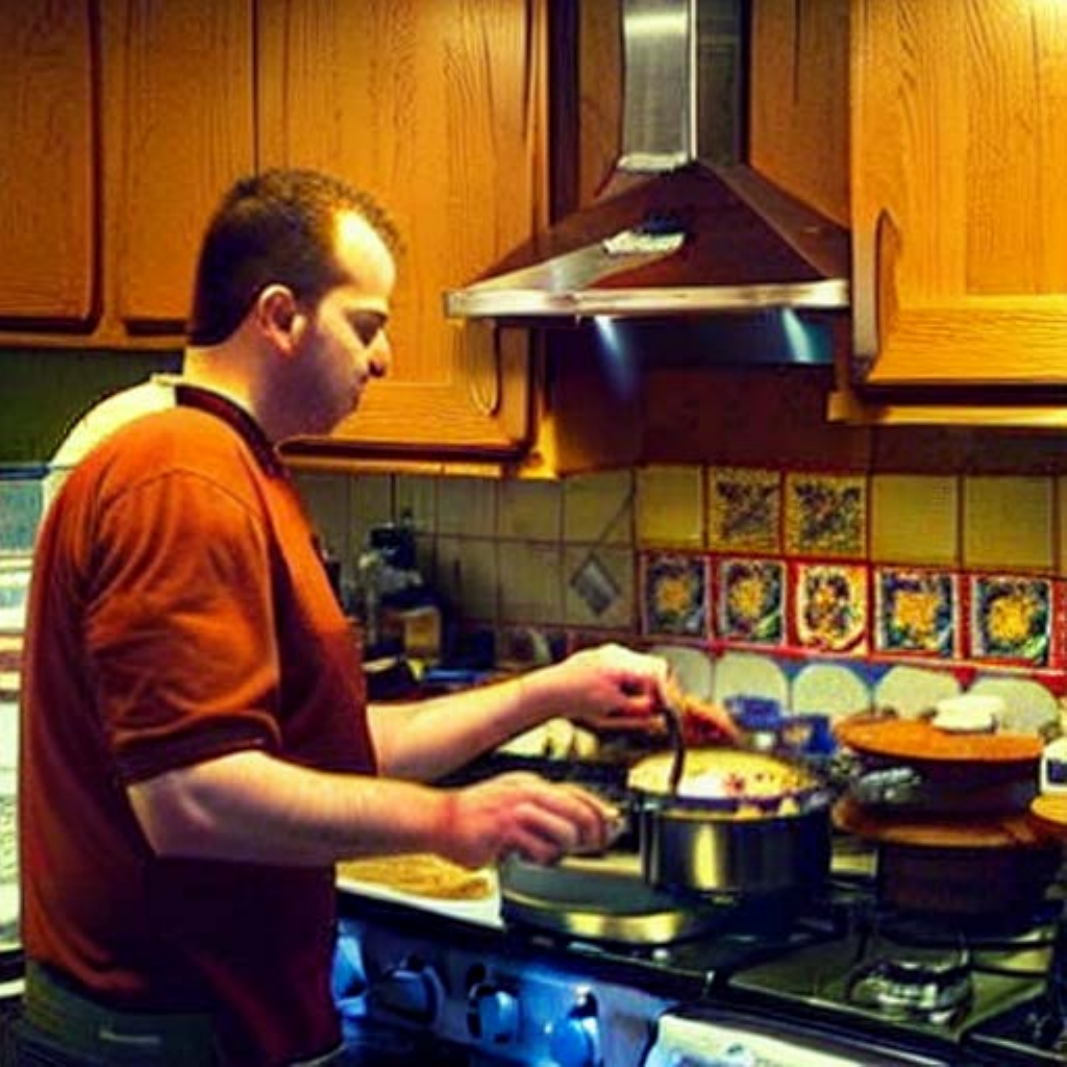} &
        \includegraphics[width=0.16\linewidth]{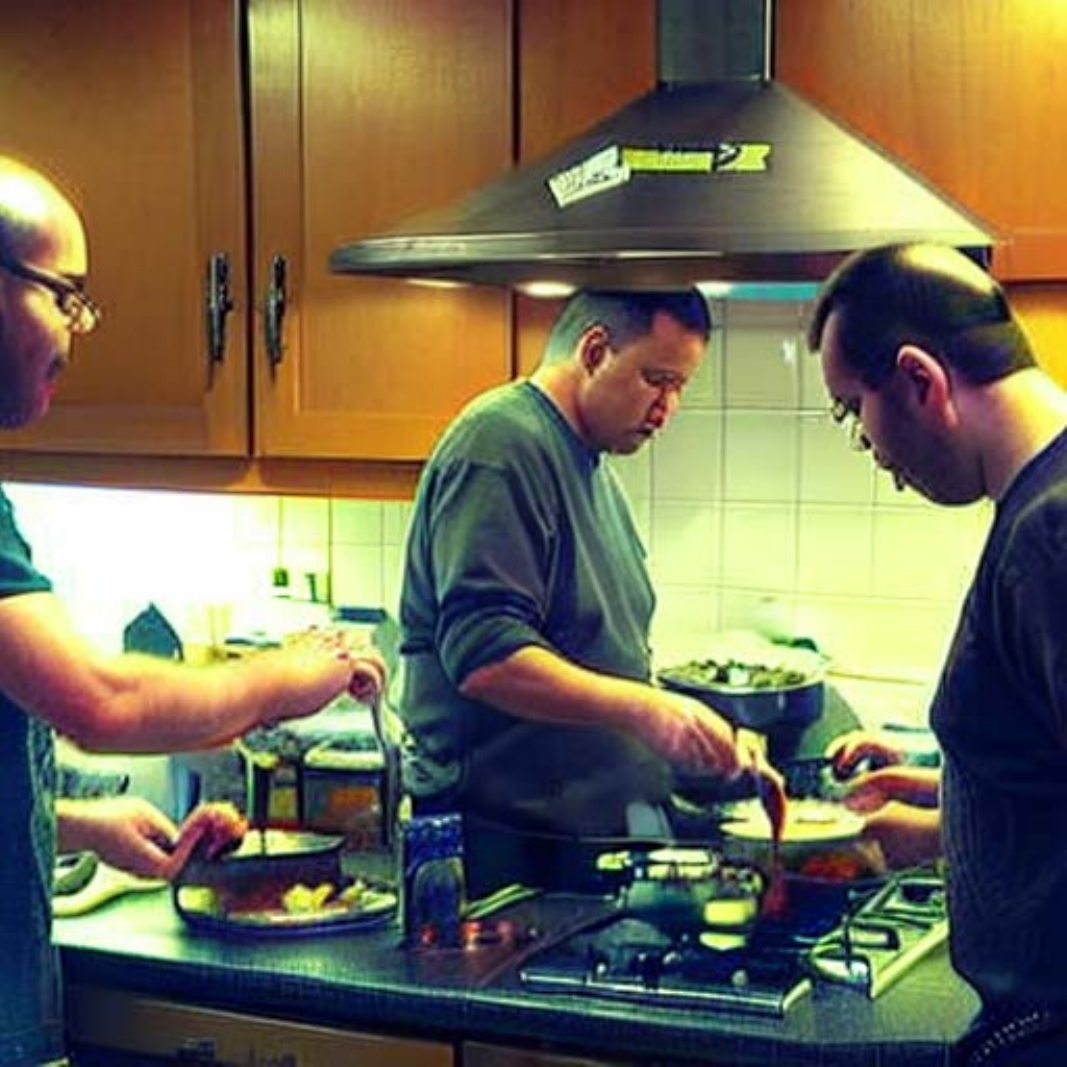} &    
        \includegraphics[width=0.16\linewidth]{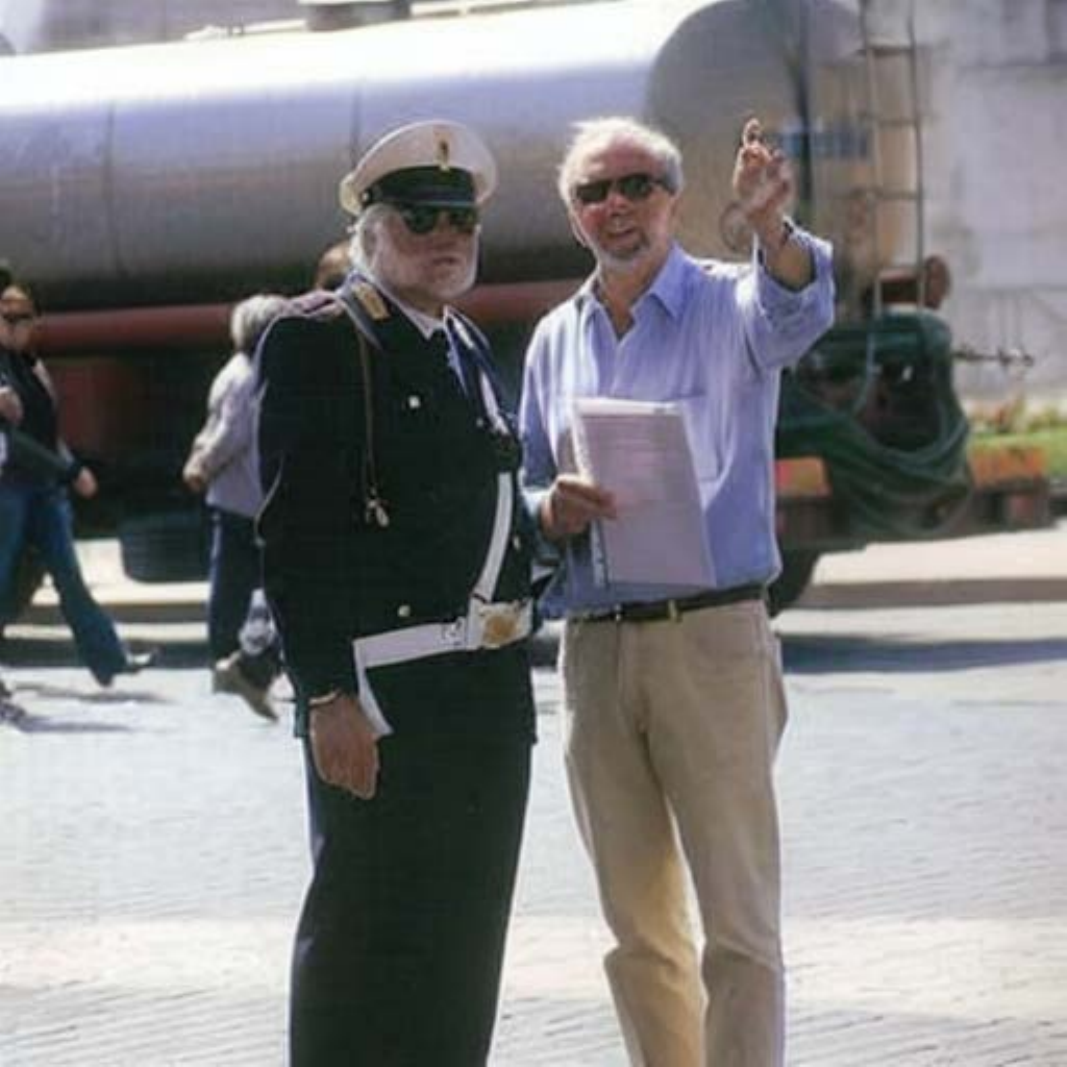} &
        \includegraphics[width=0.16\linewidth]{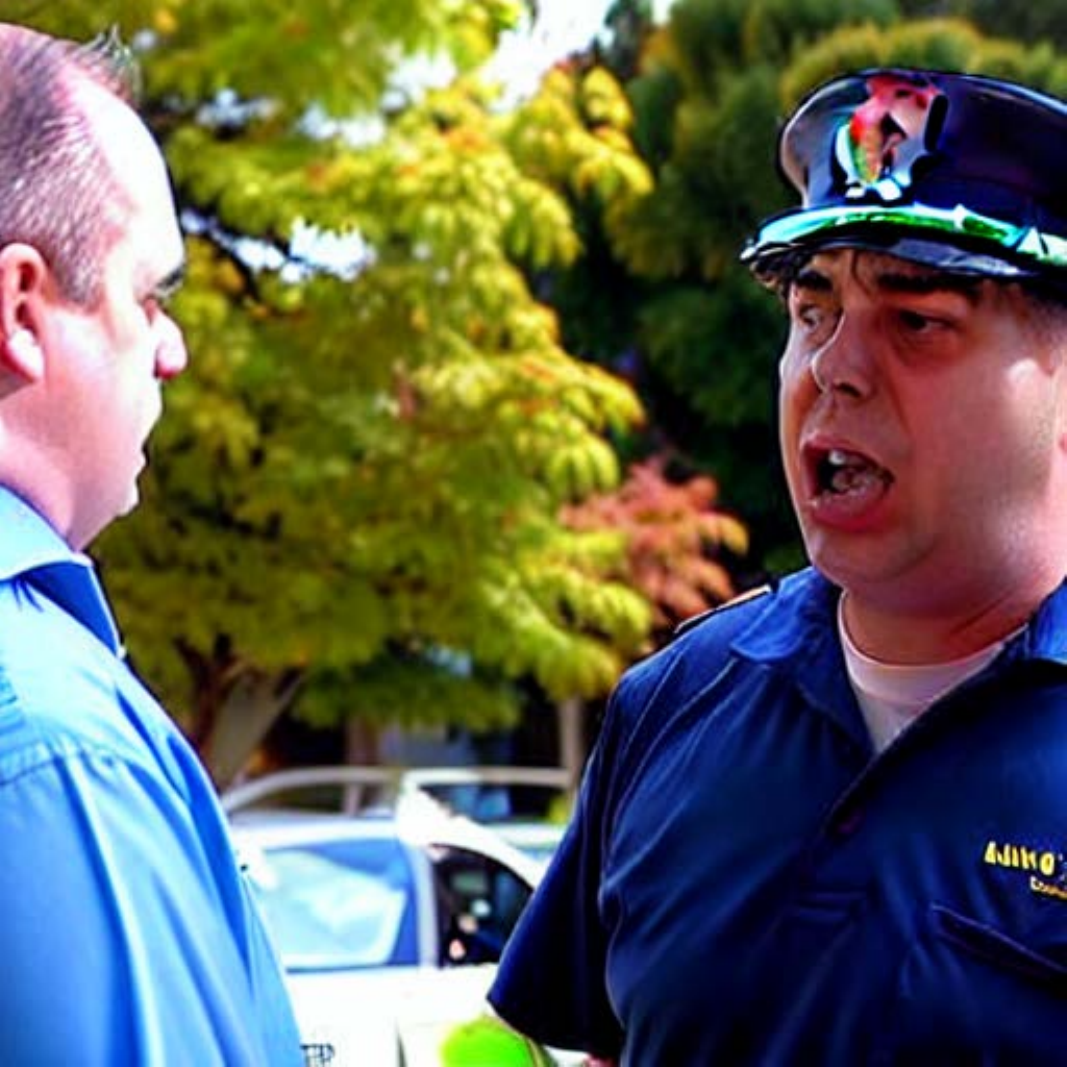} &
        \includegraphics[width=0.16\linewidth]{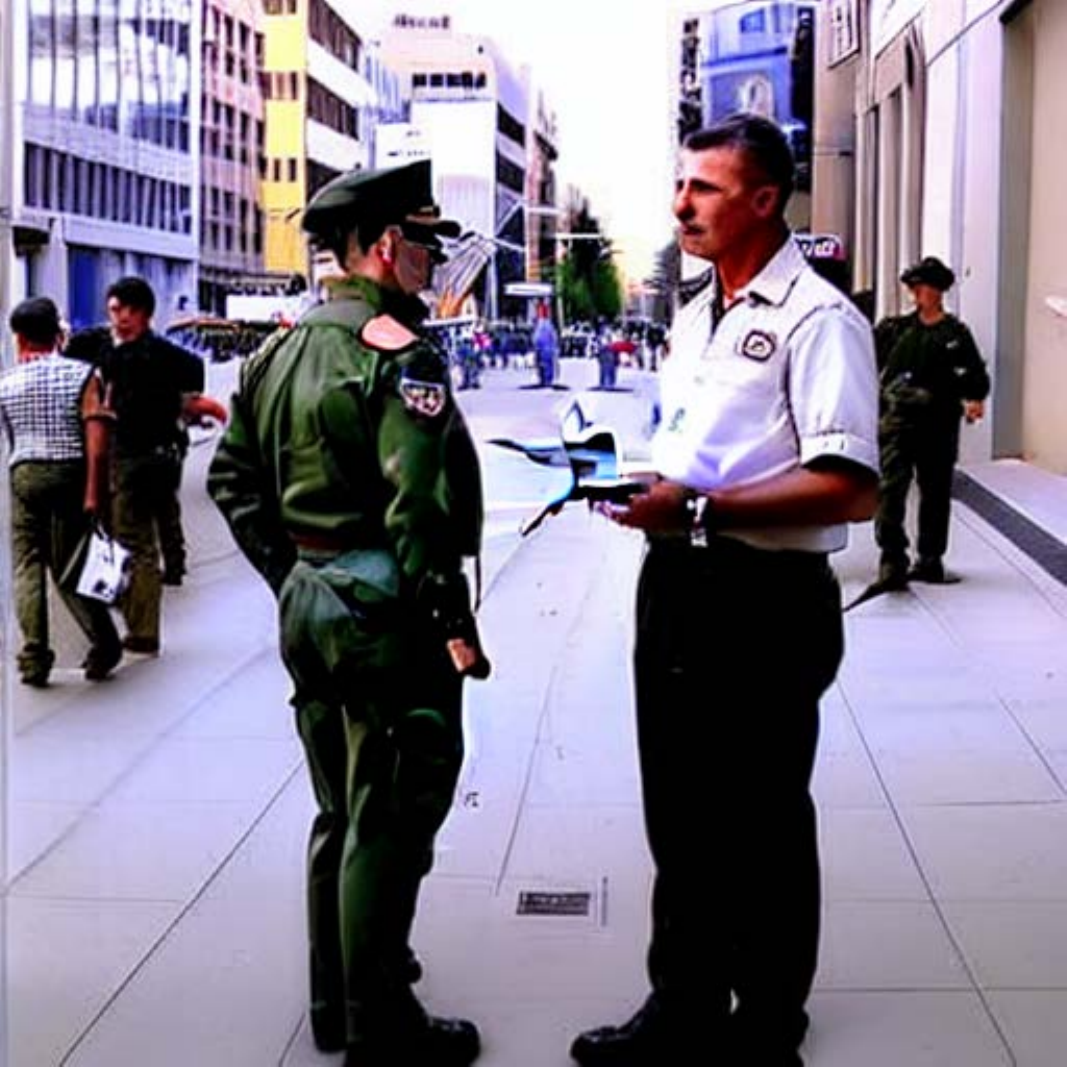} \\
    \end{tabular}
    \vspace{-5pt}
    \caption{Visualization of generation. In our method, the generated images of members are clearly closer to their originals. In Naive, both the generated member and non-member data differ significantly from their originals.}
    \label{16-flickr}
    \vspace{-5pt}
\end{figure}

We present the generation results for member and non-member samples on the Flickr dataset in Fig.~\ref{16-flickr}. The first three columns display the member images alongside their generated counterparts, obtained using random initial noise and semantic initial noise, respectively. The last three columns present non-member images and their corresponding generated results. In Naive, the generated results, whether from member or non-member samples, significantly deviate from the originals. In contrast, our method generates member images that closely resemble their originals, and the generated non-member images differ noticeably. This clear separation between member and non-member reconstruction quality is the primary reason for the attack performance achieved by our method.

\vspace{-5pt}
\section{Conclusion}
\label{conclusion}

In this paper, we reveal that standard noise schedules in diffusion models retain residual semantic information in the initial noise, leading the model to inadvertently learn correlations between the initial noise and the training images. Leveraging this vulnerability, we propose a simple yet effective membership inference attack that uses DDIM inversion to inject semantics into the initial noise and analyzes the generations. Our experiments confirm that these semantic residuals pose significant privacy risks in fine-tuned models.

\section{Acknowledgments}
This work is supported in part by New Generation Artificial Intelligence-National Science and Technology Major Project (2025ZD0123504), and in part by the Fundamental Research Funds for the Central Universities (Grant No. 2242026K30047).

\bibliographystyle{ACM-Reference-Format}
\bibliography{sample-base}





\end{document}